%% file: aanda.tex
\documentclass{aa} 
\usepackage{mathptmx}
\usepackage{txfonts}
\usepackage{float}
\usepackage{capt-of}
\usepackage{multicol}
\usepackage{newtxtext}
\usepackage[T1]{fontenc}
\usepackage{lastpage}
\DeclareRobustCommand{\VAN}[3]{#2}
\let\VANthebibliography\thebibliography
\def\thebibliography{\DeclareRobustCommand{\VAN}[3]{##3}\VANthebibliography}

\usepackage{graphicx}
\usepackage[dvipsnames]{xcolor}

\usepackage{amssymb}
\usepackage{tabularx}
\usepackage[inline]{enumitem}
\usepackage{tabulary}
\usepackage{booktabs}
\usepackage{multirow}
\usepackage{siunitx}
\usepackage{placeins}
\usepackage{listings}
\usepackage{orcidlink}
\usepackage[normalem]{ulem}
\usepackage{iftex}
\ifpdftex
  \DeclareUnicodeCharacter{2248}{$\approx$}
  \DeclareUnicodeCharacter{0306}{$\check{c}$}
\fi
\usepackage{amsmath}
\usepackage{xspace}

\usepackage{iftex}
\ifluatex
  \usepackage{fontspec}
  \newfontfamily\gurmukhifont[Path=fonts/, Extension=.ttf, Scale=0.85, RawFeature={axis={wght=300}}]{NotoSansGurmukhi-VF}
  \newfontfamily\japanesefont[Path=fonts/, Extension=.ttf, Scale=0.85, RawFeature={axis={wght=300}}]{NotoSansJP-VF}
  \newfontfamily\arabicfont[Path=fonts/, Extension=.ttf, Script=Arabic, Scale=0.85, RawFeature={axis={wght=300}}]{NotoSansArabic-VF}
  \newfontfamily\ethiopicfont[Path=fonts/, Extension=.ttf, Scale=0.85, RawFeature={axis={wght=300}}]{NotoSansEthiopic-VF}
  \newfontfamily\notofont[Path=fonts/, Extension=.ttf, Scale=0.85]{NotoSans-Light}
  \newcommand{\namegm}[1]{{\gurmukhifont #1}}
  \newcommand{\namejp}[1]{{\japanesefont #1}}
  \newcommand{\namear}[1]{{\arabicfont #1}}
  \newcommand{\nameet}[1]{{\ethiopicfont #1}}
\else
  \ifxetex
    \usepackage{fontspec}
    \newfontfamily\gurmukhifont[Path=fonts/, Extension=.ttf, Scale=0.85, RawFeature={axis={wght=300}}]{NotoSansGurmukhi-VF}
    \newfontfamily\japanesefont[Path=fonts/, Extension=.ttf, Scale=0.85, RawFeature={axis={wght=300}}]{NotoSansJP-VF}
    \newfontfamily\arabicfont[Path=fonts/, Extension=.ttf, Script=Arabic, Scale=0.85, RawFeature={axis={wght=300}}]{NotoSansArabic-VF}
    \newfontfamily\ethiopicfont[Path=fonts/, Extension=.ttf, Scale=0.85, RawFeature={axis={wght=300}}]{NotoSansEthiopic-VF}
    \newfontfamily\notofont[Path=fonts/, Extension=.ttf, Scale=0.85]{NotoSans-Light}
    \newcommand{\namegm}[1]{{\gurmukhifont #1}}
    \newcommand{\namejp}[1]{{\japanesefont #1}}
    \newcommand{\namear}[1]{{\arabicfont #1}}
    \newcommand{\nameet}[1]{{\ethiopicfont #1}}
  \else
    \newcommand{\namegm}[1]{}
    \newcommand{\namejp}[1]{}
    \newcommand{\namear}[1]{}
    \newcommand{\nameet}[1]{}
  \fi
\fi

\hypersetup{
   colorlinks=true,
   linkcolor=blue,
   citecolor=blue,
   filecolor=blue,
   urlcolor=blue
}
\newcommand\HII{\ion{H}{II}\xspace} 
\newcommand\NII{[\ion{N}{II}]\xspace} 
\newcommand\OIII{[\ion{O}{III}]\xspace} 
\newcommand\OI{[\ion{O}{I}]\xspace} 
\newcommand\SII{[\ion{S}{II}]\xspace} 

\makeatletter 
  \patchcmd{\NAT@citex}
    {\@citea\NAT@hyper@{%
      \NAT@nmfmt{\NAT@nm}%
      \hyper@natlinkbreak{\NAT@aysep\NAT@spacechar}{\@citeb\@extra@b@citeb}%
      \NAT@date}}
    {\@citea\NAT@nmfmt{\NAT@nm}%
    \NAT@aysep\NAT@spacechar\NAT@hyper@{\NAT@date}}{}{}

  \patchcmd{\NAT@citex}
    {\@citea\NAT@hyper@{%
      \NAT@nmfmt{\NAT@nm}%
      \hyper@natlinkbreak{\NAT@spacechar\NAT@@open\if*#1*\else#1\NAT@spacechar\fi}%
        {\@citeb\@extra@b@citeb}%
      \NAT@date}}
    {\@citea\NAT@nmfmt{\NAT@nm}%
    \NAT@spacechar\NAT@@open\if*#1*\else#1\NAT@spacechar\fi\NAT@hyper@{\NAT@date}}
    {}{}
\makeatother

\titlerunning{Predicting ionised gas emission in 3D with SKIRT. I.}
\title{Predicting ionised gas emission in 3D with \texttt{SKIRT}.\\ I. Framework and validation}
\author{
Anand Utsav Kapoor\inst{1} (\namegm{ਅਨੰਦ ਉਤਸਵ ਕਪੂਰ})\orcidlink{0000-0002-5187-1725}\thanks{E-mail: anandutsavkapoor@gmail.com}
\and
Maarten Baes\inst{1}\orcidlink{0000-0002-3930-2757}
\and
Aaron Smith\inst{2}\orcidlink{0000-0002-2838-9033}
\and
Arno Lauwers\inst{1}\orcidlink{0009-0003-9692-9382}
\and
Andrea Gebek\inst{1}\orcidlink{0000-0002-0206-8231}
\and
Peter Camps\inst{1}\orcidlink{0000-0002-4479-4119}
\and
Sven De Rijcke\inst{1}\orcidlink{0000-0001-7680-2059}
\and
Arjen van der Wel\inst{1}\orcidlink{0000-0002-5027-0135}
\and
Kosei Matsumoto\inst{1} (\namejp{松本 光生})\orcidlink{0000-0002-5012-6707}
\and
William McClymont\inst{3,4}\orcidlink{0009-0009-5565-3790}
}

\institute{
Department of Physics and Astronomy, Proeftuinstraat 86 N3, B-9000 Ghent, Belgium
\and
Department of Physics, The University of Texas at Dallas, Richardson, Texas 75080, USA
\and
Kavli Institute for Cosmology, University of Cambridge, Madingley Road, Cambridge CB3 0HA, UK
\and
Cavendish Laboratory, University of Cambridge, 19 JJ Thomson Avenue, Cambridge CB3 0HE, UK
}

\date{Received XXX; accepted YYY}

\begin{document}
\label{firstpage}

\abstract
{Emission lines from ionised gas are key diagnostics
of star formation, metallicity, and ionisation conditions in galaxies.
Interpreting spatially resolved observations from integral field surveys
(e.g.\ \texttt{MaNGA}, \texttt{VLT/MUSE}, \texttt{JWST/NIRSpec}) and
comparing them with hydrodynamical simulations requires three-dimensional
photoionisation models that handle realistic geometries, dust
attenuation, and synthetic instrument output.}
{We present a new photoionisation module for the Monte Carlo radiative
transfer code \texttt{SKIRT} that predicts emission-line luminosities of ionised gas
in three dimensions, combining pre-computed \texttt{Cloudy}
tables for gas temperature and opacity with a direct calculation of ion
fractions and line emissivities.}
{The module characterises the local ionising radiation field by $\log U$ and
four spectral-shape ratios across the ionising continuum (1--6\,Ryd).
Pre-computed \texttt{Cloudy} tables map these to gas temperature and
photoionisation opacities, converging through \texttt{SKIRT}'s existing
iteration cycle.
Emission-line luminosities are computed directly: an inline solver determines
ion fractions from the converged radiation field and temperature, and evaluates
emissivities from recombination coefficients and collisional excitation rates.
We validate against \texttt{Cloudy} on 60 spherical shell models and
against \texttt{COLT} on a Milky Way-analogue galaxy.}
{Across the one-dimensional benchmark grid, hydrogen recombination lines agree with
\texttt{Cloudy} to within a few per cent (H$\alpha$ median ratio 0.97),
and the forbidden lines to within $\sim$5\%, except
\SII$\lambda6717$ (median ratio 1.23), whose offset traces a temperature
overestimate near the ionisation front.
In the three-dimensional galaxy comparison, integrated luminosities
agree with \texttt{COLT} to within 18 per cent for the hydrogen lines
and 2 per cent for \NII, while \OIII\ and \SII\ are systematically
elevated by $\sim$70 and $\sim$80 per cent.
Pixel-by-pixel correlation coefficients reach
$r \geq 0.92$, with luminosity-weighted scatter of $0.14$--$0.31$\,dex,
and broadly consistent BPT line ratios.}
{The module enables self-consistent synthetic observations
in which ionised-gas emission lines, dust attenuation, and dust re-emission
are computed in a single MCRT simulation, applicable to
any hydrodynamical simulation.}



\maketitle
\input{introduction}

\input{methodology}
\input{results}

\input{conclusions}

\section{Data availability}

\texttt{DiffuseIonizedGasMix}, together with the pre-computed \texttt{Cloudy}
tables it queries, is publicly available as part of the \texttt{SKIRT} code
base\footnote{\url{https://github.com/SKIRT/SKIRT9}}. The
validation reported here is a pilot in one key respect: every test is run at
solar, spatially uniform abundances. The module itself is more general: its
inline solver treats each element individually, so gas-phase abundances can be
varied cell by cell. The multi-metallicity tables that exercise this
capability, together with their application to cosmological galaxy simulations
with per-cell, non-solar abundances read directly from the simulation snapshot,
are presented in the next paper of the series (Kapoor et al., in prep.).

\begin{acknowledgements}
AUK acknowledges support from the Belgian Federal Science Policy Office (BELSPO) via the ESA-PRODEX programme.
AS acknowledges support through JWST AR-08709.
WM thanks the Science and Technology Facilities Council (STFC) Center for Doctoral Training (CDT) in Data Intensive Science at the University of Cambridge (STFC grant number 2742968) for a PhD studentship.
We thank Rahul Kannan for providing the Milky Way simulation data used in this work,
and Ilse De Looze for access to the Margaret computing cluster.
This research made use of Astropy, matplotlib, and NumPy.
\end{acknowledgements}


\FloatBarrier
\bibliographystyle{aa}
\bibliography{bibliography}

\FloatBarrier
\input{appendix}
\end{document}

%% file: introduction.tex
\section{Introduction}
\label{sec:intro}

\begin{figure}
  \centering
  \includegraphics[width=\columnwidth]{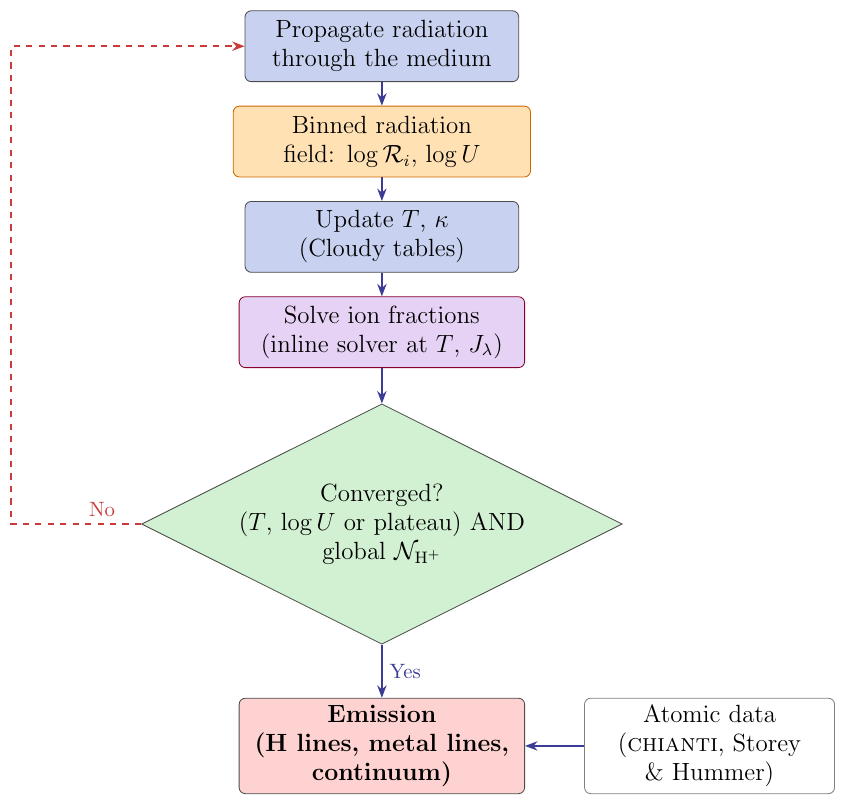}
  \caption{Overview of the per-cell state-update procedure in \texttt{DiffuseIonizedGasMix}.
           At each \texttt{SKIRT} iteration, the local ionising radiation field
           ($1$--$6$\,Ryd) is used to compute
           the ionisation parameter $\log U$ and the spectral shape ratios $R_i$
           (mean intensity ratios across five energy bins; Sect.~\ref{sec:5bin}),
           which together index pre-tabulated \texttt{Cloudy} grids to obtain the
           temperature and opacity.
           Emission-line luminosities are computed directly from the ion fractions
           determined by an inline solver at the converged temperature.
           The ionisation solver also updates the neutral fractions, and convergence is assessed
           using the criteria described in Sect.~\ref{sec:conv}.}
  \label{fig:flowchart}
\end{figure}

Emission lines from ionised gas are among the most powerful diagnostics available for
studying star-forming galaxies across cosmic time.
Hydrogen recombination lines (H$\alpha$, H$\beta$) trace star-formation rates;
forbidden metal lines (\NII, \OIII, \SII) constrain gas-phase metallicities, ionisation
conditions, electron temperatures and densities, and gas kinematics
\citep{2019ARA&A..57..511K}.
With the advent of integral field unit (IFU) spectrographs on 8-m class telescopes (VLT/MUSE,
Keck/KCWI) and, most recently, the James Webb Space Telescope (JWST)
with its NIRSpec and MIRI IFU modes, spatially resolved emission-line maps can now be
obtained for large samples of galaxies spanning a wide range of redshifts and physical
conditions \citep{2024A&A...691A..19V, 2025arXiv251016116F, 2020ARA&A..58..661F}.
Interpreting these observations reliably requires theoretical models that can predict
emission-line maps for realistic, three-dimensional galaxy geometries with comparable
resolution and physical completeness.

The ionised interstellar medium (ISM) in star-forming galaxies spans a wide
range of physical conditions.
Dense, compact \HII\ regions form around individual young stellar clusters and
dominate the total ionising photon budget, while the diffuse ionised gas (DIG)
occupies a much larger volume at lower densities
($n_{\rm e} \lesssim 10$\,cm$^{-3}$) and contributes approximately half of the
total H$\alpha$ luminosity in nearby galaxies
\citep{2016MNRAS.461.3111B, 2022A&A...659A..26B, 2022MNRAS.513.2904T, 2024MNRAS.532.2016M}.
In both regimes, the emission-line signature depends on the full
three-dimensional radiation field: \HII\ regions can overlap, share ionising
photons, and be partially shielded by intervening dust, while the DIG is
sustained primarily by Lyman-continuum photons leaking from \HII\ regions and
ionising radiation from evolved stellar populations
\citep{2022A&A...659A..26B}, though other mechanisms
(e.g.\ turbulent dissipation, cosmic rays) may also contribute
\citep{2009RvMP...81..969H}.
Predicting emission-line maps therefore requires methods that propagate the
radiation field self-consistently through the galaxy volume, accounting for
the cumulative effect of multiple stellar sources, dust attenuation, and
diffuse reemission.

Photoionisation codes such as \texttt{Cloudy} \citep{2017RMxAA..53..385F} and
\texttt{MAPPINGS} \citep{2018ascl.soft07005S} are the standard tools for
computing the emission-line spectrum of a photoionised nebula given an input radiation
field and gas geometry.
These codes solve the ionisation and thermal equilibrium with high physical fidelity,
but they operate on one-dimensional or simplified geometries and do not propagate
the radiation field self-consistently through extended, multi-source, dusty media.
At the galaxy scale, comparing simulations to observed emission-line maps therefore
typically requires either assigning each gas element a local 1D model calibrated
to a set of physical parameters, or post-processing the simulation with a dedicated
Monte Carlo radiative transfer (MCRT) code.
A complementary approach uses the on-the-fly radiation field from
radiation-hydrodynamics (RHD) simulations
\citep{2022MNRAS.511.4005K, 2025OJAp....8E.153K, 2025arXiv251005201K}
to drive per-cell photoionisation either on-the-fly or in
post-processing \citep{2018MNRAS.479..994R, 2023OJAp....6E..44K}.
However, RHD codes typically employ the M1 closure approximation
for radiation transport, which can misrepresent the radiation geometry in multi-source
environments, and numerically transient gas temperatures may propagate into
the line emissivity calculation \citep{2022MNRAS.517....1S, 2025arXiv251013952M}.

Three-dimensional Monte Carlo photoionisation codes offer a more complete treatment
of the radiation geometry.
\texttt{MOCASSIN} \citep{2003MNRAS.340.1136E} introduced fully three-dimensional
photoionisation modelling on arbitrary grids, and its dusty extension
\citep{2005MNRAS.362.1038E} couples this with dust radiative
transfer for individual nebulae.
\texttt{ART$^2$} \citep{2012MNRAS.424..884Y, 2020MNRAS.494.1919L} combines dust
continuum transfer with ionisation, Ly$\alpha$, and atomic fine-structure lines
on adaptive grids.
\texttt{CMacIonize} \citep{2018A&C....23...40V, Vandenbroucke2020, McCallum2024} and \texttt{COLT}
\citep{Smith2015, Smith2019, 2020ApJ...905...27S, 2022MNRAS.517....1S, 2025arXiv251013952M} have brought 3D photoionisation to
moving-mesh geometries and comprehensive Monte Carlo frameworks applicable to galaxy and cosmological simulation outputs.
\texttt{SIROCCO} \citep{2025MNRAS.536..879M} takes a similar Monte Carlo approach
for axially symmetric outflows in accreting systems.
By implementing photoionisation within \texttt{SKIRT}, we build on its existing
infrastructure: advanced spatial grids and optimised traversal techniques
\citep{2013A&A...554A..10S, 2014A&A...561A..77S, 2013A&A...560A..35C, 2024A&A...689A..13L},
dust models with stochastic heating \citep{2015A&A...580A..87C},
Monte Carlo optimisation techniques
\citep{2011ApJS..196...22B, 2016A&A...590A..55B, 2022A&A...666A.101B},
hybrid parallelisation \citep{2017A&C....20...16V, 2020A&C....3100381C},
kinematics \citep{2020A&C....3100381C},
and polarisation support
\citep{2017A&A...601A..92P, 2021A&A...653A..34V, 2024A&A...689A.297V}.

\texttt{SKIRT} \citep{2015A&C.....9...20C, 2020A&C....3100381C} is an open-source MCRT
code widely applied to hydrodynamical simulations to produce synthetic broadband
photometry, SEDs, and polarization maps.
Its medium framework has been extended to handle X-ray radiative transfer
\citep{2023A&A...674A.123V}, Lyman-$\alpha$ line transfer
\citep{2021ApJ...916...39C}, and non-LTE molecular line emission
\citep{2023A&A...678A.175M}, and its source framework includes
stellar spectral energy distribution (SED) families from simple stellar populations to binary-evolution libraries
such as \texttt{BPASS} \citep{2017PASA...34...58E}, as well as the
\texttt{TODDLERS} sub-grid emission library for star-forming regions
\citep{2023MNRAS.526.3871K, 2024A&A...692A..79K}.
However, until now \texttt{SKIRT} has not been able to compute the ionisation
state and emission-line luminosities of the gas on its spatial grid from the
propagated radiation field.

A partial solution was provided by the \texttt{TODDLERS} library
\citep{2023MNRAS.526.3871K, 2024A&A...692A..79K}, which couples pre-computed
\texttt{Cloudy} photoionisation models with \texttt{SKIRT} to generate emission
SEDs for individual star-forming regions as sub-grid sources.
\texttt{TODDLERS} handles the compact \HII\ region population effectively, but
treats each region in isolation: the ionising radiation from one source does not
affect the gas in adjacent cells, and the contribution of
photons escaping into the diffuse ISM is not followed on the \texttt{SKIRT} grid.

In this paper we present \texttt{DiffuseIonizedGasMix}, a new module for
\texttt{SKIRT} that enables photoionisation modelling of ionised gas on the
MCRT grid.
The module uses a hybrid approach: pre-computed \texttt{Cloudy} tables map the local
radiation field (characterised by the ionisation parameter $\log U$ and four
spectral-shape ratios) to the gas temperature and photoionisation opacities, while
emission-line luminosities are computed directly from ion fractions determined by an
inline photoionisation solver at the converged temperature.
The module is coupled to \texttt{SKIRT}'s iterative radiation propagation framework,
so that the ionisation state converges with the radiation field
as photons propagate through the galaxy volume, including scattering and absorption
by dust.
This adds photoionisation to \texttt{SKIRT}'s existing capabilities, providing a
single framework that combines emission-line predictions with dust attenuation,
dust emission, kinematics, broadband photometry, and synthetic observations for
arbitrary galaxy geometries.

We demonstrate that the approach is physically sound and numerically viable, and
characterise its accuracy across a range of physical conditions.
Broader applicability beyond the solar metallicity adopted here, including
cosmological simulations and larger galaxy samples, is left to future work.

The paper is organised as follows.
In Sect.~\ref{sec:methods} we describe the numerical implementation of
\texttt{DiffuseIonizedGasMix} within \texttt{SKIRT}.
In Sect.~\ref{sec:results} we present the results: first a validation against
\texttt{Cloudy} using 60 one-dimensional benchmark models
(Sect.~\ref{sec:results_1d}), and then a three-dimensional application to an isolated
Milky Way-analogue galaxy \citep{2020MNRAS.499.5732K}
(Sect.~\ref{sec:results_3d}).
We summarise and conclude in Sect.~\ref{sec:conclusions}.

%% file: methodology.tex

\section{The \texttt{DiffuseIonizedGasMix} module}
\label{sec:methods}
\label{sec:DIG}

\begin{figure}
  \centering
  \includegraphics[width=\columnwidth]{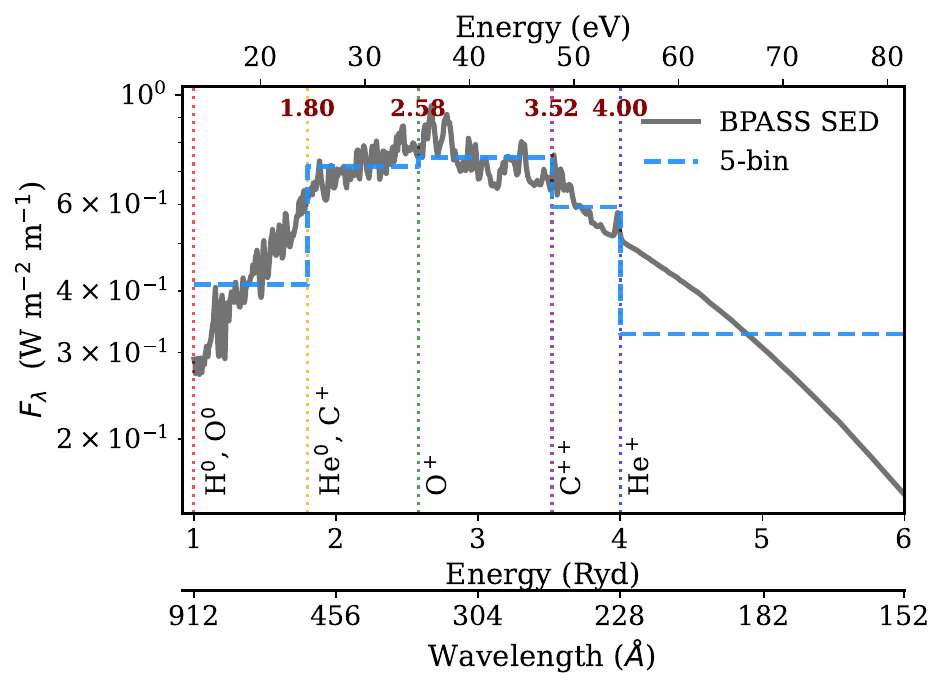}
  \caption{Five-bin structure overlaid on a \texttt{BPASS} SED
           (10\,Myr, $Z = 0.014$; black).
           The step function (blue dashed) shows the mean flux in each bin.
           Vertical dotted lines mark the bin boundaries at 1.00, 1.80, 2.58, 3.52, 4.00,
           and 6.00 Ryd, which coincide with major ionisation edges as labelled.
           This binning is designed to accurately capture the dominant
           recombination and cooling channels with minimal spectral resolution.}
  \label{fig:sed_bins}
\end{figure}

We introduce \texttt{DiffuseIonizedGasMix}\footnote{The name reflects the module's primary
application: in cosmological galaxy simulations, dense \HII\ regions are typically unresolved
and handled by subgrid prescriptions, so the resolved ionised gas that the module acts upon is
predominantly diffuse. The module itself is not restricted to low densities; its tables
extend to $n_{\rm H} = 10^3\,\rm cm^{-3}$ and can be expanded further
(Sect.~\ref{sec:density_extrap}).}, a new material-mix module in \texttt{SKIRT} that represents
photoionised gas in a self-consistent radiative-transfer framework.
The module handles emission, absorption opacities, and diffuse reemission simultaneously, and
updates its own state iteratively as the radiation field converges.
The module operates on the local radiation field and gas density in each cell,
making it agnostic to the source of the input data: it can be applied equally to
individual \HII\ regions, planetary nebulae, or galaxy-scale simulations.
In the current implementation the ionising range spans $1$--$6$\,Ryd and the
tables assume solar-scaled abundances (Sect.~\ref{sec:outlook}), but neither
choice is a fundamental restriction of the framework.

\subsection{Role in the \texttt{SKIRT} iteration cycle}

\label{sec:DIG_role}

Like other \texttt{SKIRT} medium components, the module stores a set of physical variables
per grid cell (the ionisation state, temperature, and radiation-field characterisation)
and updates them after each photon-propagation iteration based on the accumulated local
ionising radiation field ($1$--$6$\,Ryd).
The loop repeats until global convergence is declared (Sect.~\ref{sec:conv}).
Figure~\ref{fig:flowchart} gives an overview of the full per-cell update procedure.

Each cell stores per-cell state variables: the hydrogen and helium neutral
fractions ($x_{\mathrm{H}^0}$, $x_{\mathrm{He}^0}$), the ionisation parameter
($\log U$), four spectral shape ratios ($\log \mathcal{R}_2$,
$\log \mathcal{R}_3$, $\log \mathcal{R}_4$, $\log \mathcal{R}_5$; defined in
Sect.~\ref{sec:5bin}), the gas metallicity $Z$, temperature $T$, and the
ionised-hydrogen number density $n_{\mathrm{H}^+}$.
Temperatures and opacities are obtained by interpolation in pre-tabulated
\texttt{Cloudy} grids (Sect.~\ref{sec:stab}).
At each iteration, an inline ionisation solver (Sect.~\ref{sec:direct_emission})
uses the current temperature and the local wavelength-resolved radiation field
to update the hydrogen and helium neutral fractions, consistent with the ion
fractions of the heavier elements that enter the emission calculation.
Convergence is assessed jointly from three criteria described in
Sect.~\ref{sec:conv}: a per-cell criterion on the relative changes in $T$ and
$\log U$, a plateau criterion on the fraction of converged cells, and a global
criterion on the total ionised-hydrogen mass.
Emission-line luminosities are computed in the final iteration directly from
the converged ion fractions.

The combination of \texttt{SKIRT}'s Monte Carlo radiation transport with
\texttt{Cloudy}-derived tables for the thermal balance is a deliberate design
choice, rather than implementing a fully on-the-fly multi-element solver.
We note that per-iteration runtime is not the primary motivation: photon
propagation dominates the cost in any Monte Carlo photoionisation scheme, and
a full multi-element thermal solver adds only modest per-cell overhead.
The more relevant difference is in convergence behaviour: a coupled
heating-cooling solver generally requires additional iterations to settle, and
the per-cell solve must contend with multi-valued thermal equilibria in
marginally-ionised regions.
Tabulating the converged \texttt{Cloudy} steady state in advance reduces the
per-cell update to a smooth, single-valued interpolation, which removes this
source of convergence difficulty.
The tabulated approach also imports \texttt{Cloudy}'s continuously developed
and widely benchmarked microphysics, used to realise accurate \emph{local}
temperatures and opacities in each cell, without reimplementing the
heating-cooling balance in parallel.
The table interface is modular: alternative microphysics engines, expanded
element coverage, or non-solar abundance patterns can be incorporated by
regenerating the tables without touching the radiation-transport core.
The trade-off is a less granular thermal solver than a fully on-the-fly
multi-element code; we quantify the resulting accuracy against \texttt{Cloudy}
in 1D (Sect.~\ref{sec:results_1d}) and against \texttt{COLT} in 3D
(Sect.~\ref{sec:results_3d}).
This first paper of the series establishes the table-driven path; forthcoming
work will apply the framework to galaxy-evolution simulation post-processing
and develop more self-consistent treatments of the underlying photoionisation
and thermal-balance physics within \texttt{SKIRT}.

\subsection{5-bin radiation-field characterisation}
\label{sec:5bin}

At each iteration, the local radiation field in every cell must be compressed into a small set of
numbers that index the pre-computed tables.
We characterise the ionising radiation field between 1 and 6 Ryd (912--152 \AA) using five
contiguous energy bins (Table~\ref{tab:bins}).

\begin{table}
  \caption{Five-bin decomposition of the ionising radiation field.
           Each bin boundary coincides with a major ionisation edge.}
  \label{tab:bins}
  \centering
  \begin{tabular}{clccc}
    \toprule
    Bin & Edge & Range (Ryd) & Range (\AA) & Range (eV) \\
    \midrule
    1 & \ion{H}{I}   & 1.00--1.80 & 912--506 & 13.6--24.5 \\
    2 & \ion{He}{I}  & 1.80--2.58 & 506--353 & 24.5--35.1 \\
    3 & \ion{O}{II}  & 2.58--3.52 & 353--259 & 35.1--47.9 \\
    4 & \ion{C}{III} & 3.52--4.00 & 259--228 & 47.9--54.4 \\
    5 & \ion{He}{II} & 4.00--6.00 & 228--152 & 54.4--81.6 \\
    \bottomrule
  \end{tabular}
\end{table}
For each bin we compute the wavelength-averaged mean intensity $\langle J_i \rangle$ by
integrating $J_\lambda$ over the bin and dividing by the bin width.
The spectral shape is then encoded by four ratios relative to bin 1:
\begin{equation}
  \mathcal{R}_i \equiv \frac{\langle J_i \rangle}{\langle J_1 \rangle} \, , \qquad i = \{2,3,4,5\}\,.
  \label{eq:Ri}
\end{equation}
Together with the ionisation parameter $\log U$, these four ratios describe the hardness and
shape of the ionising spectrum and serve as the key interpolation axes in the tables.
Figure~\ref{fig:sed_bins} illustrates the five bins overlaid on a representative \texttt{BPASS} SED,
with bin boundaries chosen to coincide with the ionisation edges of the
principal species (\ion{H}{I} at 1\,Ryd, \ion{He}{I} at 1.8\,Ryd,
\ion{He}{II} at 4\,Ryd), supplemented by an intermediate boundary at
2.58\,Ryd (the \ion{C}{III}/\ion{N}{III} edge) and an upper cutoff at
6\,Ryd beyond which the stellar photon flux is negligible.

The ionisation parameter is computed directly from the radiation field as
\begin{equation}
  U = \frac{\phi_{\mathrm{ion}}}{n_{\mathrm{H}}\,c} \, ,
  \label{eq:logU}
\end{equation}
where $\phi_{\mathrm{ion}} = \int_{>1\,\mathrm{Ryd}} (4\pi J_\lambda \lambda / hc)\,\mathrm{d}\lambda$
is the local ionising photon flux density and $n_{\mathrm{H}}$ is the hydrogen number density.

\subsection{Pre-computed STAB tables and the dual-table system}
\label{sec:stab}

Gas temperatures and photoionisation opacities are obtained by interpolation in
pre-computed tables generated with \texttt{Cloudy} and stored in \texttt{SKIRT}'s native binary stored-table format (STAB).
Two sets of tables are used:
(\textit{i}) a 7-dimensional temperature table with axes
$(Z,\,n_{\mathrm{H}},\,\log U,\,\log \mathcal{R}_2,\,\log \mathcal{R}_3,\,\log \mathcal{R}_4,\,\log \mathcal{R}_5)$,
returning $\log T$, and
(\textit{ii}) an 8-dimensional opacity table with axes
$(\lambda,\,\log U,\,\log \mathcal{R}_2,\,\log \mathcal{R}_3,\,\log \mathcal{R}_4,\,\log \mathcal{R}_5,\,Z,\,n_{\mathrm{H}})$,
returning the total opacity $\kappa$.
Emission-line luminosities are not drawn from these tables but computed directly
from the converged state, as described in Sect.~\ref{sec:direct_emission}.
We emphasise that \texttt{Cloudy} is used solely to provide accurate \emph{local}
temperatures and opacities as a function of the local radiation field and gas
properties in each cell; the radiative transfer itself, including
geometry, attenuation, and scattering, is handled entirely by \texttt{SKIRT}.
The tabulated temperature therefore reflects the photoionisation equilibrium
of each individual gas cell, not an average over an entire ionised region.

\begin{figure}
  \centering
  \includegraphics[width=\columnwidth]{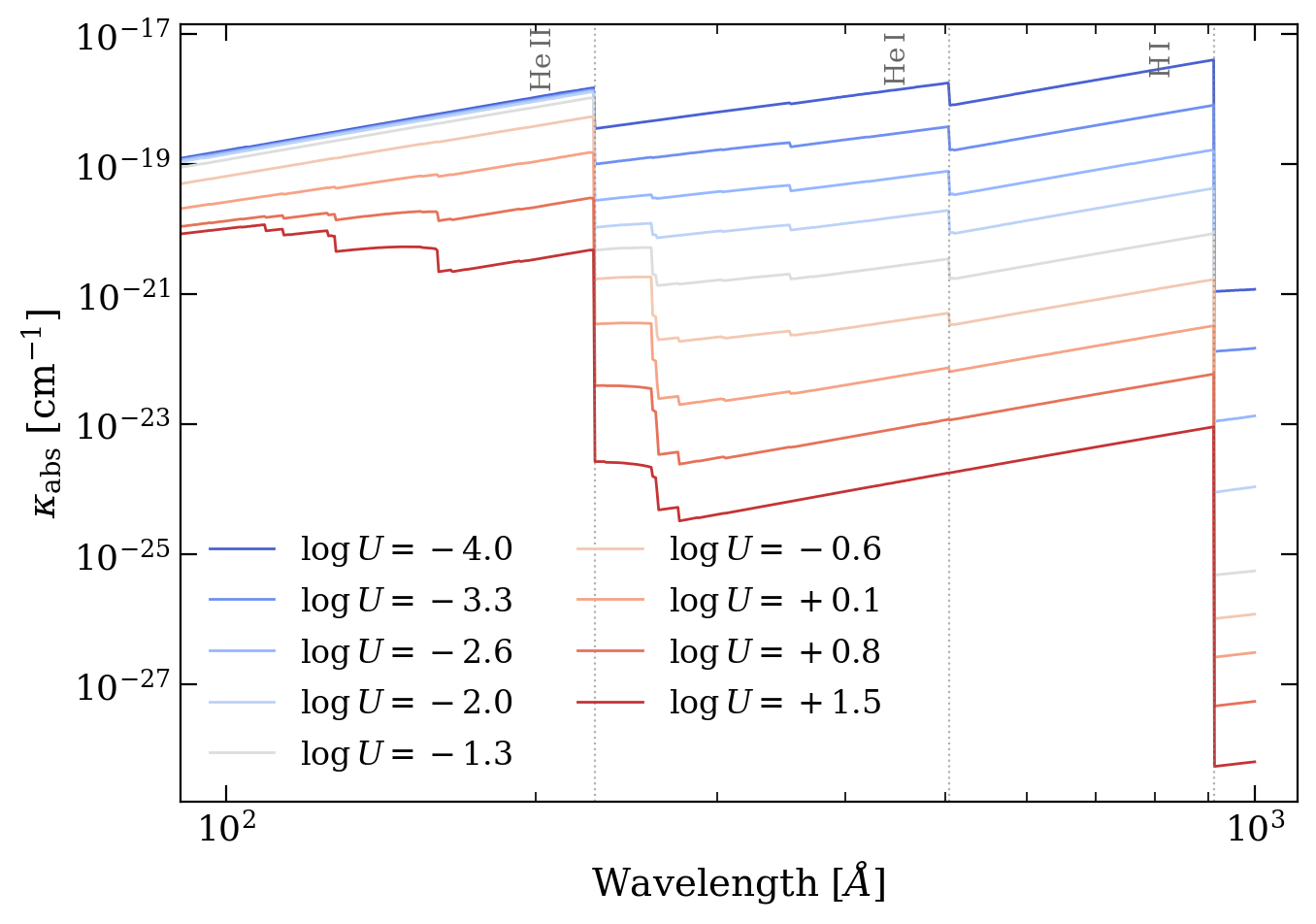}
  \caption{Absorption opacity from the standard STAB table as a function of
           wavelength for nine values of $\log U$ at fixed
           $n_{\rm H} = 10$\,cm$^{-3}$ and $\mathcal{R}_i$ fixed at the
           midpoints of their grid ranges.
           The three ionisation edges (\ion{H}{I} 912\,\AA,
           \ion{He}{I} 504\,\AA, \ion{He}{II} 228\,\AA) are clearly
           resolved despite the coarse 5-bin radiation-field characterisation.}
  \label{fig:stab_opacity}
\end{figure}

Although the radiation field is characterised by only five broad energy bins, the opacity tables are sampled at a spectral resolution
$R \equiv \lambda/\Delta\lambda \approx 300$ over the 1--6\,Ryd ionising range
and therefore resolve the full wavelength-dependent absorption structure,
including the photoionisation edges of \ion{H}{I}, \ion{He}{I}, and
\ion{He}{II}.
Figure~\ref{fig:stab_opacity} illustrates this for the standard table at
fixed $n_{\rm H} = 10$\,cm$^{-3}$ and fixed $\mathcal{R}_i$ values at the
midpoints of their respective grid ranges, showing
that the tabulated absorption cross-section varies by orders of magnitude
across the ionising continuum and responds systematically to $\log U$.

The tables are generated by running \texttt{Cloudy} \citep{2017RMxAA..53..385F} on a grid of
step-function SEDs, each constructed to match prescribed values of $\log U$ and
$(\mathcal{R}_2,\,\mathcal{R}_3,\,\mathcal{R}_4,\,\mathcal{R}_5)$.
The current tables are computed at solar metallicity ($Z = 0.014$) with
GASS10 abundances \citep{2010Ap&SS.328..179G}, which is appropriate
for all simulations used in this work (the 1D benchmark models and
the Milky Way-analogue galaxy both adopt uniform solar metallicity).
The hydrogen number density grid spans
$n_{\rm H} \in \{0.01,\,0.1,\,1,\,10,\,100,\,1000\}$\,cm$^{-3}$.
Extending the tables to a broader range of metallicities is straightforward and is planned
for future applications to cosmological simulations.

Physical conditions change rapidly near ionisation fronts, where $\log U$ drops
steeply, hydrogen transitions from fully ionised to predominantly neutral, and
the radiation field hardens due to selective absorption of soft ionising photons
\citep{2006agna.book.....O}.
A single, sparsely sampled table is insufficient to capture this regime.
We therefore maintain two separate table sets for each physical quantity.
The standard table covers $\log U \geq -4$ and is sampled at
$(-4.0,\,-3.3,\,-2.6,\,-2.0,\,-1.3,\,-0.6,\,0.1,\,0.8,\,1.5)$.
The transition table resolves the ionisation-front regime at
$\log U \leq -4$ with a dense $\log U$ grid of seven points from $-6.5$ to $-4.0$
and correspondingly expanded $\mathcal{R}_i$ coverage to capture the harder spectra
found at low ionisation parameter. The grid points are spaced non-uniformly
to cover the physically realised parameter space.
The full grid specifications for both tables are listed in Table~\ref{tab:stab_grid}.

At run time, the appropriate table is selected based on $\log U$.
In the overlap zone, the two tables (standard and transition) are blended in log-space using a linear weight:
\begin{equation}
  \log f_{\mathrm{blend}} = (1 - w)\,\log f_{\mathrm{std}} + w\,\log f_{\mathrm{trans}} \, ,
  \label{eq:blend}
\end{equation}
where $w$ increases linearly from 0 to 1 over a 0.3\,dex overlap in $\log U$.
This blending provides a smooth handoff between the two table sets,
which cannot simply be merged because they sample different $\mathcal{R}_i$ ranges
(Table~\ref{tab:stab_grid}).

\begin{table}
  \caption{Axis sampling of the standard and transition STAB tables.
           The transition table uses broader $\mathcal{R}_i$ ranges to accommodate
           the harder spectra near ionisation fronts.
           Both tables share the same $n_{\rm H}$ grid.}
  \label{tab:stab_grid}
  \centering
  \begin{tabular}{lcc}
    \toprule
    Axis & Standard & Transition \\
    \midrule
    $\log U$   & 9 pts, $[-4.0,\,+1.5]$  & 7 pts, $[-6.5,\,-4.0]$ \\
    $\log\mathcal{R}_2$ & 9 pts, $[-3.2,\,+1.2]$  & 11 pts, $[-3.3,\,+17.4]$ \\
    $\log\mathcal{R}_3$ & 9 pts, $[-4.3,\,+1.7]$  & 11 pts, $[-4.2,\,+28.6]$ \\
    $\log\mathcal{R}_4$ & 9 pts, $[-3.1,\,+1.8]$  & 11 pts, $[-3.1,\,+30.8]$ \\
    $\log\mathcal{R}_5$ & 8 pts, $[-7.8,\,+0.4]$  & 10 pts, $[-7.7,\,+31.1]$ \\
    $n_{\rm H}$ [cm$^{-3}$] & \multicolumn{2}{c}{6 pts: $0.01$--$10^3$} \\
    \bottomrule
  \end{tabular}
\end{table}

The choice of five radiation-field bins and the STAB-table node spacing
listed in Table~\ref{tab:stab_grid} is not arbitrary.
Each bin boundary coincides with a major ionisation edge
(Sect.~\ref{sec:5bin}, Table~\ref{tab:bins}), capturing the spectral
hardness changes that drive the dominant heating and cooling channels with
the smallest possible bin count.
We verified the bin-count choice empirically through a single-zone
reconstruction test (Appendix~\ref{app:bin_sensitivity},
Fig.~\ref{fig:bin_sensitivity}, Table~\ref{tab:bin_sensitivity}): at the
chosen 5-bin scheme, the equilibrium temperature and the photon-absorption
rate $\int L_{\rm full}(\nu)\,\kappa(\nu)\,\mathrm{d}\nu$ over $[1,\,6]$\,Ryd both
agree with the original to $\sim 6\%$ on average, with diminishing returns
at higher bin counts.
Refinement is costly: each additional spectral bin contributes
another spectral-ratio interpolation axis, so the table size grows
exponentially with the number of bins.
Doubling from 5 to 10 bins adds five new axes
($\mathcal{R}_6$ through $\mathcal{R}_{10}$), inflating the table by
a factor of $\sim 6 \times 10^4$ at the current node density (see
Table~\ref{tab:stab_grid}).
The corresponding accuracy gain (Appendix~\ref{app:bin_sensitivity}) is
bounded by the multi-zone 1D \texttt{Cloudy} benchmark
(Sect.~\ref{sec:results_1d}), where line luminosities across 60 models agree
with \texttt{Cloudy} at the $5$--$30\%$ level.

\subsection{Convergence criteria}
\label{sec:conv}

Convergence is declared when a per-cell or plateau criterion is satisfied
\emph{and} a global criterion on the total ionised-hydrogen mass is also
satisfied.

The per-cell criterion compares the relative change in temperature and
ionisation parameter between successive iterations.
A cell is marked as not converged if either
\begin{equation}
  \frac{\left|T^{\mathrm{new}} - T^{\mathrm{old}}\right|}{T^{\mathrm{old}}} > \epsilon_{\mathrm{cell}}
  \quad \text{or} \quad
  \frac{\left|\log U^{\mathrm{new}} - \log U^{\mathrm{old}}\right|}{\left|\log U^{\mathrm{old}}\right|} > \epsilon_{\mathrm{cell}} \, ,
  \label{eq:conv_cell}
\end{equation}
with a default threshold $\epsilon_{\mathrm{cell}} = 0.01$.
The $\log U$ check is applied only to cells above the emission floor
$\log U_{\rm emit}$ (the lower bound of the transition table; see
Sect.~\ref{sec:stab}): cells below this threshold have their temperature
pinned at the floor and produce no emission, so their $\log U$ fluctuates
from Monte Carlo noise and is not informative.
The per-cell criterion is passed when the fraction of not-converged cells
is below $10\%$ of all material-bearing cells.

A plateau criterion supplements Equation~(\ref{eq:conv_cell}): if the
converged cell fraction changes by less than $0.5\%$ across three
consecutive iterations, we accept the solution.
This prevents indefinite iteration when Monte Carlo noise keeps a small
residual population of cells from reaching the per-cell threshold.

The global criterion requires
\begin{equation}
  \frac{\left|\mathcal{N}_{\mathrm{H}^+}^{\mathrm{new}} - \mathcal{N}_{\mathrm{H}^+}^{\mathrm{old}}\right|}{\mathcal{N}_{\mathrm{H}^+}^{\mathrm{old}}} \leq \epsilon_{\mathrm{global}} \, ,
  \label{eq:conv_global}
\end{equation}
where $\mathcal{N}_{\mathrm{H}^+} = \sum_{\mathrm{cells}} n_{\mathrm{H}^+}\,V_{\mathrm{cell}}$
and the default threshold is $\epsilon_{\mathrm{global}} = 10^{-3}$.
This global gate prevents false convergence in situations where trivially
neutral cells dominate the per-cell count while the ionised structure
continues to evolve.

Table~\ref{tab:module_params} collects the default convergence parameter
values for \texttt{DiffuseIonizedGasMix}; all remaining choices (spatial grid,
wavelength grid, number of photon packets) are set through the standard
\texttt{SKIRT} configuration.
The convergence behaviour for the Milky Way application is shown in
Appendix~\ref{app:convergence}.

\begin{table}
  \caption{Default parameter values adopted for the \texttt{DiffuseIonizedGasMix} module
           in this work.}
  \label{tab:module_params}
  \centering
  \small
  \setlength{\tabcolsep}{4pt}
  \begin{tabular}{lll}
    \toprule
    Parameter & Symbol & Value \\
    \midrule
    Per-cell threshold ($T$, $\log U$) & $\epsilon_{\rm cell}$    & 0.01 \\
    Max.\ non-converged fraction       & --                       & 10\% \\
    Plateau tolerance (3 iter.)        & --                       & 0.5\% \\
    Global threshold                   & $\epsilon_{\rm global}$  & $10^{-3}$ \\
    \bottomrule
  \end{tabular}
\end{table}

\subsection{Diffuse reemission}
\label{sec:reemission}

Ionising photons absorbed by neutral gas are not simply destroyed; a fraction are re-emitted as
new ionising photons through recombination.
We implement this diffuse field as a scattering process following \citet{2004MNRAS.348.1337W}
as realized in \texttt{CMacIonize} \citep{2018A&C....23...40V}.

When an ionising photon is absorbed, it may be re-emitted into one of the following
channels: (\textit{i}) the hydrogen Lyman continuum,
(\textit{ii}) the helium Lyman continuum, (\textit{iii}) the He 19.8\,eV line from the $2^3\!S$ level, or
(\textit{iv}) the helium two-photon continuum from the $2^1\!S$ level.
For hydrogen, only recombinations to the ground state (rate coefficient $\alpha_1$) produce
a new ionising photon; recombinations to excited states yield non-ionising radiation
(Balmer series, etc.).
Channel probabilities are determined from recombination rate coefficients at the local gas
temperature, following equations (24) and (25) of \citet{2004MNRAS.348.1337W}.

When a photon is absorbed, it is assigned to H or He with probability proportional to
$n_{\mathrm{H}^0}\,\sigma_{\mathrm{H}}(\lambda)$ and $n_{\mathrm{He}^0}\,\sigma_{\mathrm{He}}(\lambda)$,
respectively.
He Ly$\alpha$ photons ($2^1\!P$) are handled via the on-the-spot (OTS) approximation: they are absorbed locally
by neutral hydrogen with probability $p_{\mathrm{OTS}}$ and either re-emitted in the hydrogen
Lyman continuum or converted to the helium two-photon continuum.
For the two-photon channel, only 56\% of events produce a photon energetic enough to
re-ionise hydrogen.
The wavelength of each re-emitted photon is drawn from temperature-dependent
cumulative distribution functions (CDFs) in $\lambda$, stored in a dedicated
three-dimensional table with axes (channel, $T$, $\lambda$).
The channel axis indexes the three emission channels (H Lyman continuum, He Lyman continuum,
He two-photon continuum); the temperature axis spans 51 logarithmically spaced points from
100 to $10^5$\,K; and the wavelength axis contains 1001 points covering the respective
continuum ranges.
The H and He Lyman continuum spectra follow the Wood--Mathis--Ercolano form
$J(\nu) \propto \nu^3\,\sigma(\nu)\,\exp[-h(\nu-\nu_0)/kT]$ with photoionisation
cross-sections from \citet{1996ApJ...465..487V}, while the He two-photon continuum is sampled
from the tabulated spectral shape of \citet{1969PhRv..180...25D}.
All reemission is isotropic and unpolarized.

Implementing the reemission as scattering rather than local absorption ensures that
the diffuse ionising field produced by recombinations propagates through the grid and is
included in the radiation field seen by neighbouring cells.

\subsection{Density extrapolation}
\label{sec:density_extrap}

Hydrodynamical simulations can produce gas densities outside the range covered by the STAB tables.
For the opacity tables, we apply a linear scaling
$\kappa \propto n_{\mathrm{H}}$ for cells below the table minimum, reflecting the proportionality
between column density and optical depth.
For cells above the table maximum, we clamp to the boundary value.
For the emission calculation (Sect.~\ref{sec:direct_emission}), we cap the hydrogen number
density at $n_{\mathrm{H}} = 10^3$\,cm$^{-3}$, consistent with the upper boundary of the
STAB tables.
In practice, cells exceeding this density are rare in the resolved ISM of
galaxy-scale simulations and typically correspond to compact star-forming
regions that are better treated by sub-grid prescriptions
(e.g.\ \texttt{TODDLERS}, \citealt{2024A&A...692A..79K}).
Extending the tables to higher densities is straightforward but not
required for the applications presented here.

\subsection{Emission calculation}
\label{sec:direct_emission}

Once the radiation field and gas temperature have converged through the iteration
cycle described above, we compute emission-line luminosities from the
local physical conditions in each cell rather than interpolating in pre-computed
emission tables.

The motivation for this hybrid split is that opacity, temperature, and emission
have very different sensitivities to the spectral resolution of the radiation
field characterisation.
The photoionisation opacity is dominated by hydrogen and helium, whose
ionisation edges coincide with the bin boundaries (1.0\,Ryd for H, 1.8\,Ryd
for \ion{He}{I}); moreover, the cross-sections scale as $\nu^{-3}$, so
the photoionisation rates are strongly weighted toward photons near the edge
and are therefore well captured by the bin-integrated fluxes.
The gas temperature is set by the balance between photoionisation heating
and line cooling, with the latter scaling as $\exp(-E/kT)$; this exponential
dependence acts as a thermostat, making the equilibrium temperature
insensitive to modest errors in the heating rate.
This argument holds when photoionisation dominates the heating budget;
in regions where shocks or supernova feedback contribute significantly
to the thermal energy, the equilibrium temperature may depend on
additional heating channels not captured by the tables
\citep[e.g.][]{2025arXiv251013952M}.
Emission-line luminosities, by contrast, depend on the ion fractions of
individual metal species (e.g.\ O$^{2+}$, N$^+$, S$^+$), whose ionisation
edges (35.1, 29.6, 23.3\,eV) are not all resolved by the five-bin
characterisation: the O$^{2+}$ edge at 35.1\,eV coincides with the
bin~2/3 boundary, but the N$^+$ and S$^+$ edges fall within bins~2 and~1,
respectively.
While the five-bin representation captures the overall spectral hardness
well enough for temperature and opacity, computing emission lines from
the full wavelength-resolved radiation field avoids the residual ambiguity
in metal ion fractions that the coarse binning would introduce.
By using the table-driven approach for temperature and opacity (where the
coarse spectral binning is sufficient; see Fig.~\ref{fig:stab_opacity}) and the
full wavelength-resolved radiation field for the ion fractions that enter the
emission calculation, the hybrid design reserves coarse spectral resolution for
the quantities that tolerate it and fine resolution for those that require it.

For each cell, we solve the ionisation balance of 10 elements
(H, He, C, N, O, Ne, Mg, Si, S, Fe; 84 ionisation stages in total) at the
converged temperature $T$ and local radiation field $J_\lambda$, using
photoionisation cross-sections from \citet{1996ApJ...465..487V} and recombination
coefficients from \citet{1996ApJS..103..467V}.

We compute 20 emission lines: 9 hydrogen recombination lines and 11 forbidden
metal lines.
All 84 ionisation stages are needed to determine the ion fractions of the
carrier species accurately.
The highest ionisation stages of S, Mg, Si, Ne, and Fe are truncated relative
to a fully stripped count because they are populated only by photons above
the $1$--$6$\,Ryd range tracked here and do not contribute to the optical
lines of interest; per-cell ion fractions are computed inline rather than
persisted as state, so any future extension to additional stages incurs no
storage penalty.
The 20 lines represent the brightest optical and near-infrared diagnostics
commonly used in studies of star-forming galaxies.
Adding further lines (e.g.\ far-infrared fine-structure transitions such as
\OIII$\,88\,\mu$m) requires only the corresponding collisional excitation rate ($q_{\rm col}$) tables and
is planned for a future extension.
For the hydrogen recombination lines (Ly$\alpha$ through Br$\alpha$), we evaluate
\begin{equation}\label{eq:Hline}
    L_{\rm line} = h\nu_{\rm line} \int \alpha_{\rm eff}(T, n_e)\,n_e\,n_{\rm H^+}\,\mathrm{d}V \, ,
\end{equation}
where $\alpha_{\rm eff}(T, n_e) = P_B(T, n_e)\,\alpha_B(T)$ is the effective
recombination coefficient for the specific transition.
We adopt Case~B recombination coefficients from \citet{1997MNRAS.292...27H} and
line emission probabilities $P_B$ (the probability that a given recombination
produces a photon in the specific transition) from \citet{1995MNRAS.272...41S}.
We do not include collisional excitation of the hydrogen lines; this
contribution is negligible for the Balmer lines at typical \ion{H}{II}-region
temperatures ($T \lesssim 2 \times 10^4$\,K) but can become significant
for Ly$\alpha$ at higher temperatures
\citep[see appendix of][]{2022MNRAS.517....1S} and will be added in a
future update.

For the 11 forbidden metal lines (\NII$\lambda\lambda6548,6583$;
\OI$\lambda\lambda6300,6364$; [\ion{O}{II}]$\lambda\lambda3726,3729$;
\OIII$\lambda\lambda4363,4959,5007$; \SII$\lambda\lambda6717,6731$), we evaluate
\begin{equation}\label{eq:metalline}
    L_{\rm line} = h\nu_{\rm line} \int q_{\rm col}(T, n_e)\,n_e\,n_{\rm ion}\,\mathrm{d}V \, ,
\end{equation}
where $n_{\rm ion}$ is the number density of the carrier ion (obtained from the
solved ionisation fractions and the element abundance) and
$q_{\rm col}(T, n_e)$ is the collisional excitation rate coefficient.
We pre-tabulate $q_{\rm col}$ from the \texttt{CHIANTI} atomic database
\citep{1997A&AS..125..149D, 2021ApJ...909...38D} by solving the statistical
equilibrium for each carrier ion on a grid in $(T, n_e)$.

In addition to the line emission, we compute the nebular continuum
(free-bound, free-free, and two-photon emission from H and He) using the
formalism of \citet{2006MNRAS.372.1875E}.

%% file: results.tex
\section{Results}
\label{sec:results}

\begin{figure}[htbp]
  \centering
  \includegraphics[width=\columnwidth]{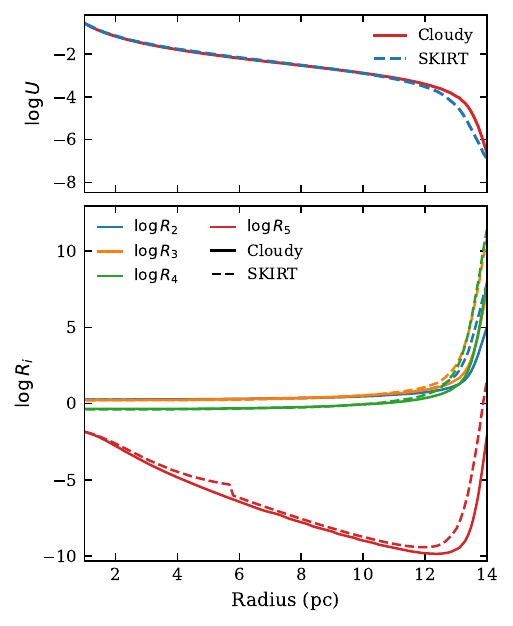}
  \caption{Radiation field characterisation as a function of radius for the
           representative benchmark model ($Q_{\rm ion} = 10^{49}$\,s$^{-1}$,
           $n_{\rm H} = 10$\,cm$^{-3}$, age 10\,Myr).
           \textit{Top:} ionisation parameter $\log U$.
           \textit{Bottom:} spectral-shape ratios
           $\log\mathcal{R}_2$--$\log\mathcal{R}_5$ (colours as labelled).
           Solid lines are \texttt{Cloudy}; dashed lines are \texttt{SKIRT}.
           The two codes agree to within $\sim$0.1\,dex across the ionised volume;
           the largest deviations appear near the ionisation front.}
  \label{fig:radiation_profile}
\end{figure}

We validate \texttt{DiffuseIonizedGasMix} in two stages.
First, we compare against \texttt{Cloudy} in a controlled one-dimensional geometry over a grid of
physical conditions (Sect.~\ref{sec:results_1d}).
Second, we apply the module to a cosmological galaxy simulation and compare against the
\texttt{COLT} Monte Carlo photoionisation code (Sect.~\ref{sec:results_3d}).
In both cases we focus on the photoionisation physics alone: dust is not
included in any of the validation runs, so that differences between codes
can be attributed entirely to the photoionisation modelling.

\subsection{One-dimensional benchmark against \texttt{Cloudy}}
\label{sec:results_1d}

\subsubsection{Setup}
\label{sec:results_1d_setup}

We construct a suite of 60 one-dimensional test models spanning a grid of five ionising
photon luminosities $Q_{\rm ion} \in \{10^{48},\,3{\times}10^{48},\,10^{49},\,3{\times}10^{49},\,10^{50}\}$\,s$^{-1}$,
three hydrogen number densities $n_{\rm H} \in \{1,\,10,\,100\}$\,cm$^{-3}$, and four \texttt{BPASS}
stellar population ages (1\,Myr, 10\,Myr, 100\,Myr, 1\,Gyr).
All models assume solar metallicity ($Z = 0.014$).
Since $\log U$ is set by $Q_{\rm ion}$ and $n_{\rm H}$ alone (age only affects the
spectral shape at fixed $Q_{\rm ion}$), the 15 $(Q_{\rm ion},\,n_{\rm H})$ combinations
span nine distinct ionisation parameters listed in Table~\ref{tab:logU_grid}, ranging from
$\log U \approx -2.6$ to $+1.5$.
These values are measured at the inner (illuminated) face of the shell; within the nebula,
$\log U$ declines as ionising photons are absorbed, so each model traverses a range of
ionisation conditions from the bright inner zone down to low-ionisation gas approaching the
ionisation front.

\begin{figure}[htbp]
  \centering
  \includegraphics[width=\columnwidth]{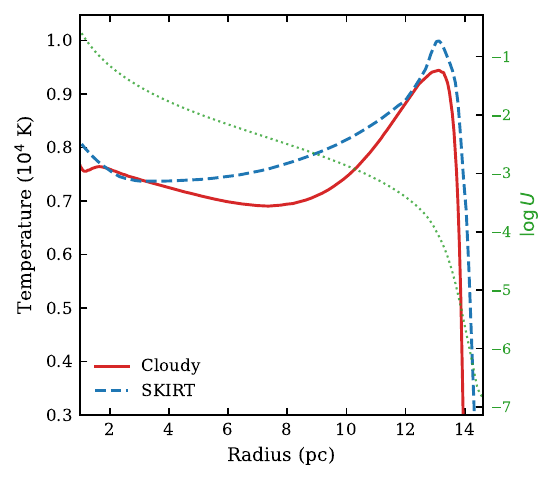}
  \caption{Gas temperature as a function of radius for the representative
           benchmark model ($Q_{\rm ion} = 10^{49}$\,s$^{-1}$,
           $n_{\rm H} = 10$\,cm$^{-3}$, age 10\,Myr, $\log U \approx -0.6$).
           The \texttt{Cloudy} reference (solid red) and the \texttt{SKIRT}
           radial-mean profile (dashed blue) agree to within a few per cent
           across the ionised zone, with larger deviations near the ionisation
           front.}
  \label{fig:temp_profile}
\end{figure}

\begin{table}
  \caption{Ionisation parameter $\log U$ at the inner boundary for each combination of
           $Q_{\rm ion}$ and $n_{\rm H}$ in the benchmark grid (solar metallicity,
           $Z = 0.014$), read from the \texttt{Cloudy} output files.
           Column headers show $\log_{10}(Q_{\rm ion}/\mathrm{s}^{-1})$; values 48.5 and
           49.5 correspond to $3{\times}10^{48}$ and $3{\times}10^{49}$\,s$^{-1}$,
           respectively.
           Each of the 15 $(Q_{\rm ion}, n_{\rm H})$ combinations is run at four
           stellar ages, giving 60 models in total.}
  \label{tab:logU_grid}
  \centering
  \begin{tabular}{lccccc}
    \toprule
    & \multicolumn{5}{c}{$\log(Q_{\rm ion}/\mathrm{s}^{-1})$} \\
    $n_{\rm H}$ [cm$^{-3}$] & 48.0 & 48.5 & 49.0 & 49.5 & 50.0 \\
    \midrule
    1   & $-0.55$ & $-0.07$ & $\phantom{-}0.45$ & $\phantom{-}0.93$ & $\phantom{-}1.45$ \\
    10  & $-1.55$ & $-1.07$ & $-0.55$           & $-0.07$           & $\phantom{-}0.45$ \\
    100 & $-2.55$ & $-2.07$ & $-1.55$           & $-1.07$           & $-0.55$           \\
    \bottomrule
  \end{tabular}
\end{table}

To provide an independent reference, we also run the same 60 models with \texttt{COLT} on a
one-dimensional spherical grid.
For each model, the shell edges are taken directly from the
\texttt{Cloudy} zone boundaries (including the adaptive refinement \texttt{Cloudy} applies
near the ionisation front), so that the radial resolution matches between the two setups.
An additional 20 logarithmically-spaced shells are appended from the outermost \texttt{Cloudy}
zone to the domain boundary to cover the neutral gas beyond the ionisation front.
The inner radius is fixed at 1\,pc and GASS10 solar abundances are adopted throughout.
The ionisation equilibrium is solved with $5 \times 10^6$ photon packets, followed by a dedicated
line-transfer step with the same number of packets per line for each of the five benchmark lines.
The total line luminosity is extracted directly from the converged solution and compared
against both \texttt{Cloudy} and \texttt{SKIRT}.

For each model, \texttt{SKIRT} simulates a uniform-density spherical shell irradiated by
the \texttt{BPASS} spectral energy distribution at the corresponding age.
The shell is discretised on a one-dimensional spherical grid with 1000
linearly spaced radial bins, well beyond the $\sim$50 bins at which the
line luminosities are converged to within 2\%.
The inner radius is fixed at 1\,pc; the outer radius is taken from the
\texttt{Cloudy} output for each model, ensuring that both codes model
the same physical volume so that any differences reflect the physics
rather than the geometry.
The ionising radiation field is stored on a linear wavelength grid of 51
points between 1 and 6\,Ryd, which convergence tests show is sufficient
(all lines within 6\% of a 501-point grid).
The one-dimensional grid concentrates all cells along the radial
direction, so Monte Carlo noise is low even with $10^5$ photon packets per
iteration; each model runs for a minimum of five iterations.
The same incident SED and gas column density are used to run \texttt{Cloudy}.

We compare the predicted luminosities of five optical emission lines:
H$\alpha$\,$\lambda6563$, H$\beta$\,$\lambda4861$,
\NII$\lambda6583$, \OIII$\lambda5007$, and \SII$\lambda6717$.
These lines are drawn from different ionisation zones within the nebula:
the hydrogen recombination lines are produced throughout the ionised volume,
\OIII\ traces the high-ionisation inner region,
and \NII\ and \SII\ probe the outer, lower-ionisation gas approaching the
ionisation front \citep{2019ARA&A..57..511K}.
Together they provide a test of the module across the full depth
of the modelled shell.
Line luminosities are computed by the inline solver
(Sect.~\ref{sec:direct_emission}) and compared against the corresponding
\texttt{Cloudy} intensities.

\subsubsection{Radiation field characterisation}
\label{sec:results_1d_rad}

Before comparing temperatures and emission lines, we examine how faithfully
\texttt{DiffuseIonizedGasMix} captures the local radiation field as it propagates
through the shell.
The module characterises the field at each cell by the ionisation parameter $\log U$
and four spectral-shape ratios $\log\mathcal{R}_2$--$\log\mathcal{R}_5$,
defined as the ratio of the mean intensity in each of the four higher-energy bins
($1.8$--$2.58$, $2.58$--$3.52$, $3.52$--$4.0$, and $4.0$--$6.0$\,Ryd)
to that in the lowest bin ($1.0$--$1.8$\,Ryd; see Sect.~\ref{sec:methods}).
Figure~\ref{fig:radiation_profile} shows the radial profiles of these five
quantities for the representative model ($Q_{\rm ion} = 10^{49}$\,s$^{-1}$,
$n_{\rm H} = 10$\,cm$^{-3}$, age 10\,Myr), comparing \texttt{SKIRT} against the
zone-by-zone output of \texttt{Cloudy}.

\begin{figure}[htbp]
  \centering
  \includegraphics[width=\columnwidth]{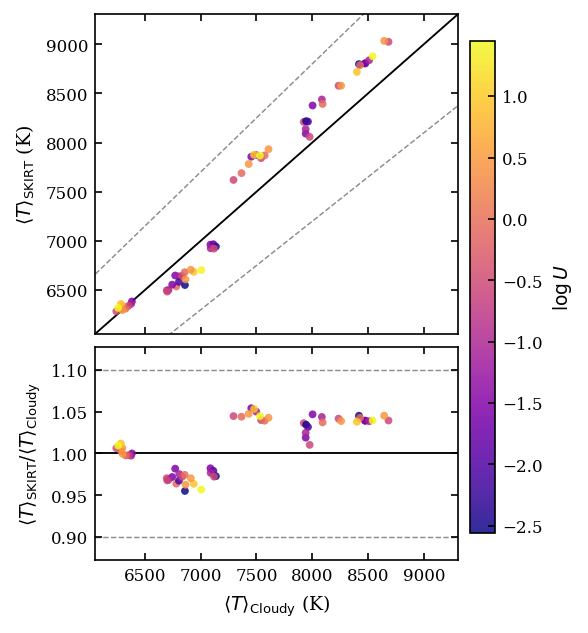}
  \caption{Temperature comparison across all 60 benchmark models.
           Ratio of mean \texttt{SKIRT} to mean \texttt{Cloudy} temperature versus
           $\langle T \rangle_{\rm Cloudy}$, coloured by $\log U$.
           The solid line marks perfect agreement and dashed lines
           show $\pm$5\%.
           The median ratio is $1.020$ with an $L_1$ error of 2.6\%.}
  \label{fig:temp_summary}
\end{figure}

The $\log U$ profiles agree throughout the ionised zone, with both codes
reproducing the characteristic decline driven by geometric dilution and absorption.
The spectral-shape ratios $\log\mathcal{R}_i$ reflect the progressive hardening
of the radiation field as softer photons are preferentially absorbed at smaller radii.
\texttt{SKIRT} reproduces the \texttt{Cloudy} $\mathcal{R}_i$ profiles to within
$\sim$0.1\,dex across the bulk of the ionised volume; the largest deviations appear
near the ionisation front, where the radiation field hardens rapidly as soft
photons are exhausted; in this narrow zone the five-bin spectral
characterisation is the main limiting factor, as the rapid spectral
hardening cannot be fully captured by so few radiation-field bins.
Table interpolation errors may also contribute.

\subsubsection{Temperature structure}
\label{sec:results_1d_temp}

Figure~\ref{fig:temp_profile} shows the gas temperature as a function of radius
for the same representative model.
The \texttt{SKIRT} profile agrees well with \texttt{Cloudy} throughout the
ionised region, reproducing both the roughly flat plateau at
$T \approx 7500$\,K in the inner zone and the temperature rise approaching
the ionisation front.
Small residuals reflect the finite resolution of the STAB interpolation tables.

Figure~\ref{fig:temp_summary} extends the comparison across all 60 models.
For each model, the mean temperature $\langle T \rangle$ is computed as the
unweighted average over 100 uniformly spaced radial points between the inner
radius and the Str\"omgren radius $r_{\rm S}$, determined independently for
each code as the radius where $x_{\rm H^0} = 0.5$.
The two codes predict similar Str\"omgren radii across the benchmark grid.
The median \texttt{SKIRT}-to-\texttt{Cloudy} temperature ratio is $1.020$
with a mean absolute relative error ($L_1$) of 2.6\%.
Individual models span the range $0.961$--$1.046$, with no clear
trend with ionisation parameter or stellar spectrum, confirming that the
STAB temperature tables perform well across the full parameter space
explored here.

\subsubsection{Emission line comparison}
\label{sec:results_1d_lines}

\begin{figure}[htbp]
  \centering
  \includegraphics[width=\columnwidth]{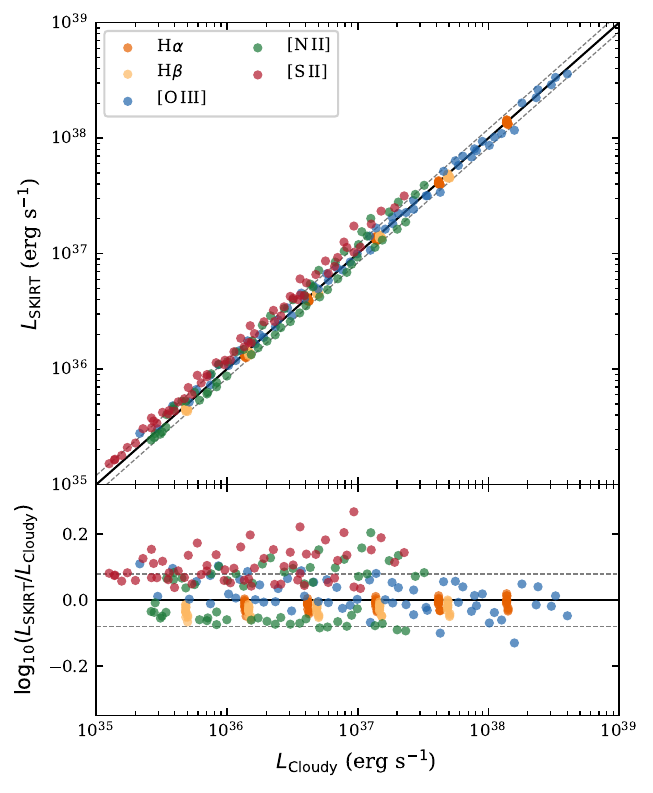}
  \caption{Comparison of \texttt{SKIRT} and \texttt{Cloudy} emission-line
           luminosities across all 60 benchmark models.
           \textit{Top:} \texttt{SKIRT} versus \texttt{Cloudy} luminosity for
           all five lines (colours as labelled); the solid diagonal marks
           perfect agreement and dashed lines show factors of $1.2$ and $1/1.2$.
           \textit{Bottom:} $\log_{10}(L_{\rm SKIRT}/L_{\rm Cloudy})$
           as a function of $L_{\rm Cloudy}$; dashed lines mark
           $\pm\log_{10}(1.2) \approx \pm 0.08$\,dex.
           H$\alpha$ and H$\beta$ agree with \texttt{Cloudy} to within
           a few per cent in the median (ratios 0.97 and 0.94);
           \SII\ shows the largest offset (median ratio 1.23).}
  \label{fig:line_summary}
\end{figure}

Figure~\ref{fig:line_summary} compares \texttt{SKIRT} and \texttt{Cloudy}
luminosities for all five emission lines across the 60 benchmark models;
Table~\ref{tab:line_stats} summarises the median ratio and mean absolute
relative error (L1) for each line.
The upper panel shows the direct scatter comparison; the lower panel shows
the ratio $L_{\rm SKIRT}/L_{\rm Cloudy}$ as a function of the \texttt{Cloudy}
luminosity.

The hydrogen recombination lines are well reproduced, with median
$L_{\rm SKIRT}/L_{\rm Cloudy}$ ratios of $0.97$ for H$\alpha$ and $0.94$
for H$\beta$.
The forbidden metal lines \OIII\ and \NII\ show median ratios of $1.02$
and $1.04$, respectively.
\SII\ has a median ratio of $1.23$, the largest offset among the five lines.

\begin{figure}[htbp]
  \centering
  \includegraphics[width=\columnwidth]{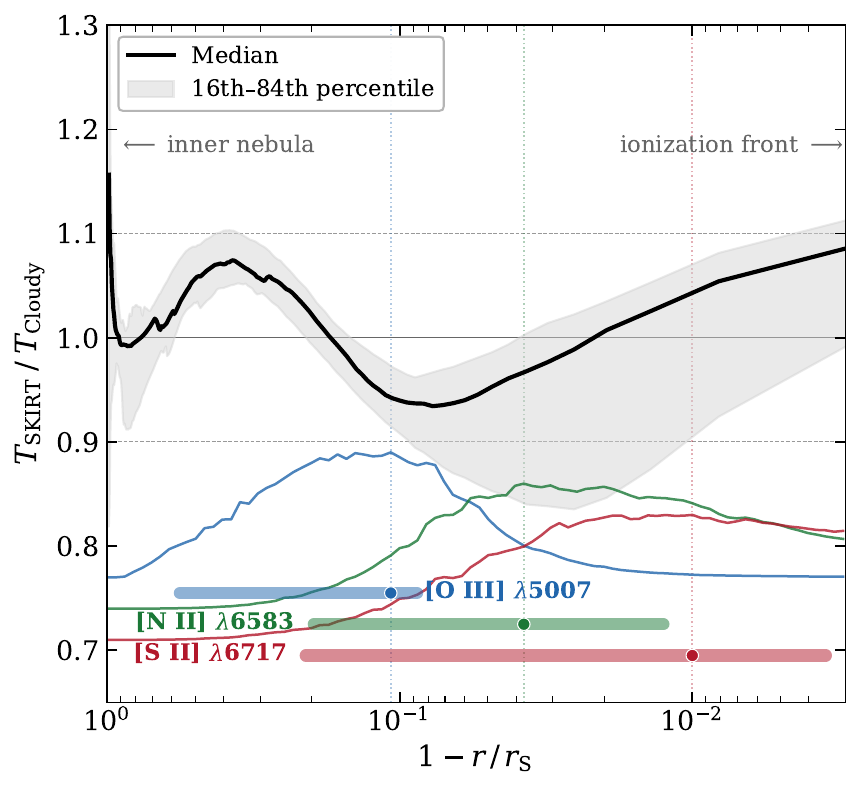}
  \caption{Median $T_{\rm SKIRT}/T_{\rm Cloudy}$ versus $1 - r/r_{\rm S}$
           across all 60 benchmark models (black curve; grey band: 16th--84th
           percentile).
           Coloured bars show the radial range producing 80\% of each line's
           luminosity (from \texttt{Cloudy}); dots mark the peak of
           $\mathrm{d}L/\mathrm{d}r$, the differential luminosity per
           unit radius.
           Thin coloured curves show the median
           $\mathrm{d}L/\mathrm{d}r$ profile (normalised to unit peak)
           for each line.
           \OIII\ is produced in the interior where the radiation field
           varies slowly (Fig.~\ref{fig:radiation_profile}), while \NII\
           and \SII\ peak near the ionisation front where the radiation
           field changes rapidly and the temperature is overestimated
           by up to $\sim$10\%.
           The logarithmic $1 - r/r_{\rm S}$ axis resolves the thin
           ionisation-front region where conditions change most rapidly.}
  \label{fig:temp_ratio_profile}
\end{figure}

\begin{figure*}[tbp]
  \centering
  \includegraphics[width=\textwidth]{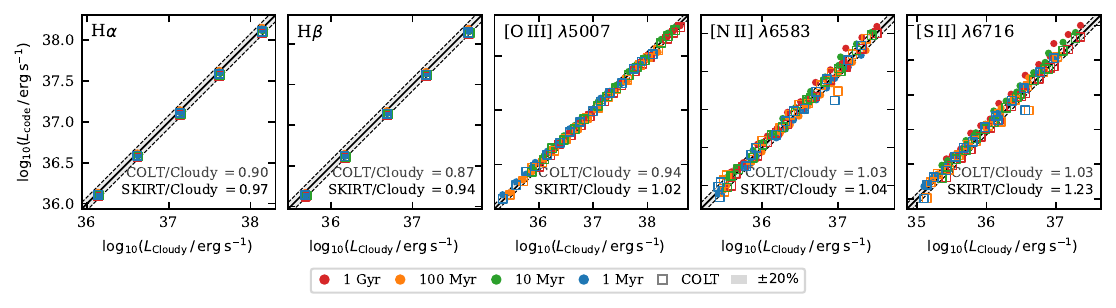}
  \caption{Emission-line luminosities from \texttt{SKIRT} (filled circles) and
           \texttt{COLT} (open squares) versus \texttt{Cloudy} across the 60
           one-dimensional benchmark models, coloured by stellar age.
           The solid diagonal marks perfect agreement; dashed lines and the grey
           band indicate $\pm$20\%.
           Median ratios relative to \texttt{Cloudy} are quoted in each panel;
           both codes reproduce the \texttt{Cloudy} luminosities to within
           $\sim$10\% in the median, except \SII\ for \texttt{SKIRT}
           (median ratio 1.23).}
  \label{fig:1d_colt}
\end{figure*}

Figure~\ref{fig:temp_ratio_profile} explains this pattern.
It shows how the temperature residual varies with radius, plotted against
$1 - r/r_{\rm S}$ on a logarithmic axis to resolve the ionisation-front region.
Each code is normalised by its own Str\"omgren radius, determined numerically
as the radius where the neutral hydrogen fraction crosses $x_{\rm H^0} = 0.5$;
the two codes predict consistent Str\"omgren radii across the benchmark grid.
The temperature agrees to within a few per cent in the nebular interior but is
overestimated by up to $\sim$10\% near the ionisation front, where the radiation
field changes rapidly over a small radial interval.
The coloured bars show where each forbidden line is predominantly produced
($10^{\rm th}$--$90^{\rm th}$ percentile of the \texttt{Cloudy} cumulative emissivity; dots mark
peak emissivity), following the classical ionisation stratification
\citep{2019ARA&A..57..511K}: \OIII\ throughout the interior (peak at
$0.86\,r/r_{\rm S}$), \NII\ near the front ($0.97\,r/r_{\rm S}$), and \SII\
at the very edge ($0.99\,r/r_{\rm S}$).
The ion fractions that enter the forbidden-line luminosities are solved
from the radiation field at the converged temperature (Sect.~\ref{sec:direct_emission}),
so a temperature residual does propagate into the ion balance through the
recombination and collisional-ionisation rate coefficients.
However, these rate coefficients vary more gradually with $T$ than the
collisional excitation rates that set the line emissivities, which depend
exponentially on temperature through $\exp(-E/kT)$.
A temperature overestimate therefore affects the line luminosities
primarily through enhanced excitation, with the ion-fraction
shift playing a secondary role.
This is negligible for \OIII, which is produced in the interior where the
radiation field varies slowly, but accounts for much of the \SII\ offset.

Figure~\ref{fig:1d_colt} extends the comparison to include \texttt{COLT}, showing
\texttt{SKIRT} (filled circles) and \texttt{COLT} (open squares) luminosities against
\texttt{Cloudy} for all five lines, coloured by stellar age.
Both codes recover the \texttt{Cloudy} Balmer luminosities to within $\sim$10\%
in the median.
For the metal lines, \texttt{COLT} shows median ratios of $\approx 1.03$ for \NII\
and $\approx 0.94$ for \OIII, while \texttt{SKIRT} obtains $1.04$ for \NII\
and $1.02$ for \OIII.
For \SII, \texttt{COLT} obtains a median ratio of $1.03$, compared to $1.23$ for \texttt{SKIRT}.
Both codes thus reproduce the forbidden metal lines to within $\sim$10\% of
\texttt{Cloudy} in the median, with \SII\ being the main outlier for \texttt{SKIRT}.

\begin{table}
  \caption{Summary of emission-line agreement between \texttt{SKIRT} and
           \texttt{Cloudy} across the 60 benchmark models.
           The median ratio refers to $L_{\rm SKIRT}/L_{\rm Cloudy}$;
           the L1 error is the mean absolute relative error,
           $\langle |r_i - 1| \rangle$, where $r_i$ is the ratio for each model.}
  \label{tab:line_stats}
  \centering
  \begin{tabular}{llcc}
    \toprule
    Line & Zone & Median ratio & L1 error \\
    \midrule
    H$\alpha$\,$\lambda6563$ & throughout      & $0.97$ & $3.5\%$ \\
    H$\beta$\,$\lambda4861$  & throughout      & $0.94$ & $6.5\%$ \\
    \OIII\,$\lambda5007$     & inner           & $1.02$ & $9.6\%$ \\
    \NII\,$\lambda6583$      & near front      & $1.04$ & $18\%$ \\
    \SII\,$\lambda6717$      & at front        & $1.23$ & $28\%$ \\
    \bottomrule
  \end{tabular}
\end{table}

Taken together, the 1D benchmark demonstrates that \texttt{DiffuseIonizedGasMix}
reproduces the temperature structure and emission-line luminosities of
photoionised nebulae across a wide range of physical conditions.
The hydrogen recombination lines agree with \texttt{Cloudy} to within
2--6\% in the median, and the forbidden lines \OIII\ and \NII\ to within
2--4\%.
\SII\ shows a larger median offset of 23\%, driven by the temperature
overestimate near the ionisation front where \SII\ is produced
(Fig.~\ref{fig:temp_ratio_profile}).
These results provide a basis for the three-dimensional application in
Sect.~\ref{sec:results_3d}.

\subsection{Three-dimensional application}
\label{sec:results_3d}

\begin{figure*}[tbp]
  \centering
  \includegraphics[width=0.85\textwidth]{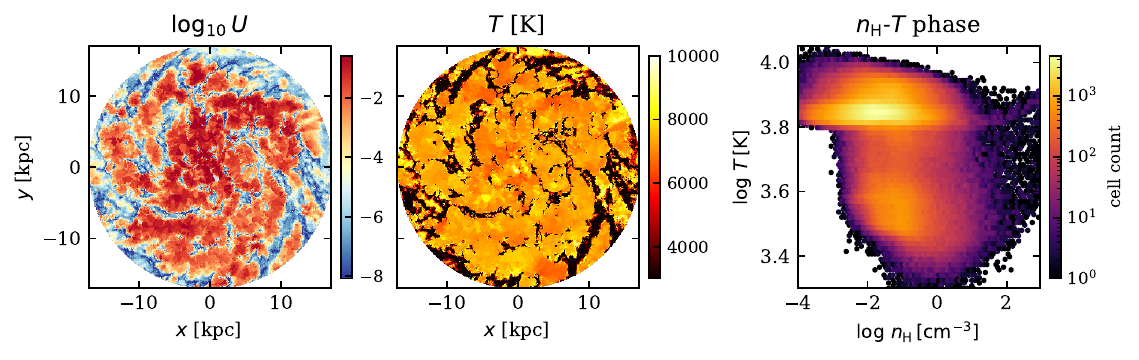}
  \caption{Physical conditions of the ionised gas in the
           \citet{2020MNRAS.499.5732K} galaxy as inferred by \texttt{SKIRT}.
           Left: ionisation parameter $\log U$; centre: gas temperature $T$.
           Both maps show a face-on midplane cut through the disc,
           projected onto a $1024 \times 1024$ pixel grid.
           Right: $n_{\rm H}$--$T$ phase diagram constructed from the
           per-cell distribution of all ionised cells in the analysis
           volume, colour-coded by cell count.
           The bulk of the ionised gas occupies a narrow band centred
           on $T \approx 7000$\,K extending across roughly six decades
           in density.}
  \label{fig:3d_physical}
\end{figure*}

\subsubsection{Simulation setup}
\label{sec:results_3d_setup}

We apply \texttt{DiffuseIonizedGasMix} to the gas and stellar distribution of an
isolated Milky Way-analogue simulation from \citet{2020MNRAS.499.5732K}, run with the
\texttt{SMUGGLE} stellar feedback framework \citep{2019MNRAS.489.4233M} within the
moving-mesh code \texttt{Arepo} \citep{2010MNRAS.401..791S}, and including
radiation hydrodynamics \citep{2019MNRAS.485..117K} and non-equilibrium thermochemistry.
The snapshot provides gas densities, temperatures, metallicities, and stellar
particle properties, which are post-processed with \texttt{COLT} \citep{2025arXiv251013952M}
and with \texttt{SKIRT} independently.
Both codes use the same \texttt{BPASS} stellar SEDs with a
\citet{2003PASP..115..763C} IMF and an upper stellar mass limit of $100\,\rm M_\odot$,
the same GASS10 solar abundances, and compare projected emission-line maps for
H$\alpha$, H$\beta$, \OIII$\lambda5007$, \NII$\lambda6583$, and \SII$\lambda6717$ at four inclinations
($0^\circ$, $30^\circ$, $60^\circ$, $90^\circ$).
Both codes discretise the galaxy volume onto the native Voronoi mesh
imported from the \texttt{Arepo} snapshot.
To ensure a consistent comparison, both codes apply a density ceiling of
$n_{\rm H} = 10^3\,\rm cm^{-3}$ (Sect.~\ref{sec:density_extrap}):
in \texttt{SKIRT} this is the upper boundary of the STAB tables, and
\texttt{COLT} was explicitly configured to match.
Table~\ref{tab:mw_config} summarises the key \texttt{SKIRT} configuration
parameters for this run.

\begin{table}
  \caption{\texttt{SKIRT} configuration for the Milky Way-analogue simulation.}
  \label{tab:mw_config}
  \centering
  \small
  \begin{tabular}{@{}ll@{}}
    \toprule
    Parameter & Value \\
    \midrule
    \multicolumn{2}{l}{\textit{Source}} \\
    SED family        & \texttt{BPASS} v2.2, Chabrier $100\,\rm M_\odot$ \\
    Source wavelength range & $0.014$--$0.091\,\mu$m \\
    Smoothing kernel  & Cubic spline \\[2pt]
    \multicolumn{2}{l}{\textit{Medium}} \\
    Spatial grid      & Voronoi \\
    Domain            & $(\pm 30\,\rm kpc)^3$ \\
    Abundances        & GASS10 solar ($Z = 0.014$, uniform) \\
    Density ceiling   & $n_{\rm H} = 10^3\,\rm cm^{-3}$ \\[2pt]
    \multicolumn{2}{l}{\textit{Radiation field}} \\
    RF wavelength grid & Linear, 102 pts, $0.01$--$0.091\,\mu$m \\[2pt]
    \multicolumn{2}{l}{\textit{Iteration}} \\
    Photon packets    & $10^9$ (full iteration) \\
    Packet ramp       & start $1\%$, factor $1.3$ per iteration \\[2pt]
    \multicolumn{2}{l}{\textit{Convergence}} \\
    Per-cell threshold ($T$ and $\log U$) & 0.01 \\
    Max unconverged fraction & 10\% \\
    Stability plateau tolerance & 0.5\% \\
    Global $\Delta n_{\rm H^+}$ threshold & 0.1\% \\[2pt]
    \multicolumn{2}{l}{\textit{Instrument}} \\
    Output wavelength grid & $R = 3000$, $0.45$--$0.70\,\mu$m \\
    Field of view     & $40 \times 40$\,kpc ($400 \times 400$\,px) \\
    Distance          & 10\,Mpc \\
    \bottomrule
  \end{tabular}
\end{table}

\subsubsection{Galaxy-scale ionised gas comparison}
\label{sec:results_3d_gal}

The gas distribution spans conditions from dense \HII\ regions to the more diffuse
ionised ISM.
\texttt{SKIRT} propagates radiation from all stellar particles through the galaxy-wide
gas volume; \texttt{COLT} provides reference
intrinsic emission-line maps from the same snapshot.

Figure~\ref{fig:3d_physical} shows the physical conditions inferred by
\texttt{SKIRT} from the converged radiation field, projected onto a
face-on midplane cut through the disc ($1024 \times 1024$ pixel grid,
where each pixel averages over the Voronoi cells within its footprint).
The ionisation parameter map reveals high-$U$ compact \HII\ regions
embedded in a diffuse, lower-$U$ medium tracing the inter-arm gas.
The temperature map is nearly uniform at $T \approx 7000$\,K across
the ionised volume, as expected for solar-metallicity photoionised gas.
The accompanying $n_{\rm H}$--$T$ phase diagram resolves two
physically distinct branches in the ionised gas.
The dominant one, at $\log T \approx 3.85$, corresponds to the
fully-ionised warm photoionised medium ($x_{\rm HII} \approx 1$)
and extends across roughly six decades in density.
A second branch at $\log T \approx 3.4$--$3.7$ traces gas at low
ionisation parameter ($\log U \lesssim -4$), where partial
ionisation ($x_{\rm HII} \approx 0.1$--$0.5$) is maintained primarily
by the diffuse Lyman continuum (LyC) field reaching
otherwise-shielded regions.
Cells fully shielded from ionising radiation settle at the 300\,K
temperature floor and fall outside the plotted range.
These converged conditions drive the emission computed in the
final iteration.

\begin{figure*}[tbp]
  \centering
  \includegraphics[width=0.85\textwidth]{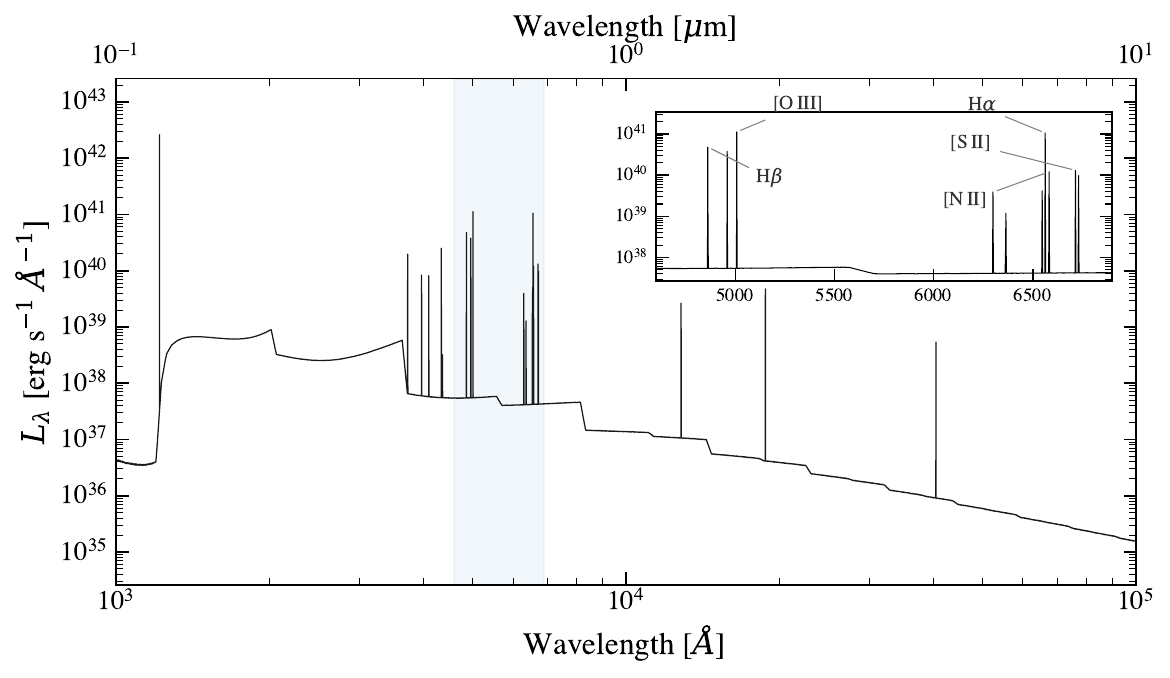}
  \caption{Integrated gas emission spectrum of the
           \citet{2020MNRAS.499.5732K} Milky Way-analogue galaxy at face-on
           inclination ($i = 0^\circ$), as produced by \texttt{SKIRT}.
           The spectrum contains all 20 emission lines currently implemented
           (Sect.~\ref{sec:direct_emission}); the line list is extensible
           by supplying the appropriate atomic data.
           The inset zooms into the optical window ($0.46$--$0.69\,\mu$m),
           labelling the principal emission lines; H$\alpha$ dominates the
           optical emission.}
  \label{fig:sed_mw}
\end{figure*}

Figure~\ref{fig:sed_mw} shows the integrated spectral energy distribution
of the galaxy as produced by \texttt{SKIRT}, from the ionising UV through the
near-infrared.
The spectrum contains all 20 emission lines computed by the module;
the optical inset labels the five lines whose maps are compared below.

\begin{figure*}[tbp]
  \centering
  \includegraphics[width=\textwidth]{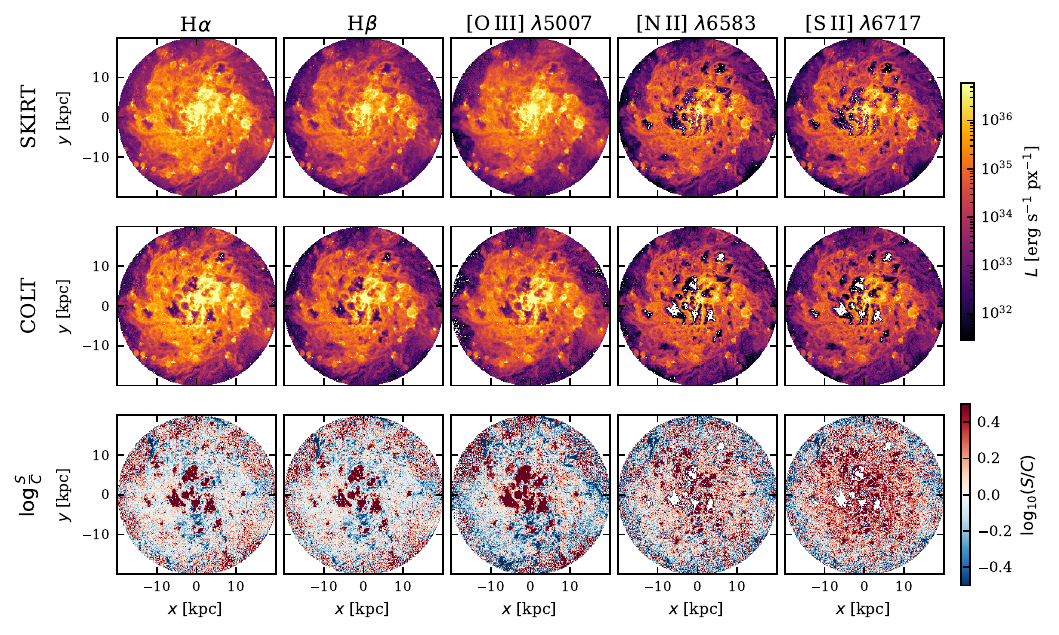}
  \caption{Face-on ($i = 0^\circ$) emission-line maps of the
           \citet{2020MNRAS.499.5732K} Milky Way-analogue galaxy.
           Top row: \texttt{SKIRT} (\texttt{DiffuseIonizedGasMix});
           middle row: \texttt{COLT} reference; bottom row:
           $\log_{10}(\mathrm{SKIRT}/\mathrm{COLT})$ residual.
           From left to right: H$\alpha$\,$\lambda6563$,
           H$\beta$\,$\lambda4861$, \OIII$\lambda5007$,
           \NII$\lambda6583$, and \SII$\lambda6717$.
           The top two rows share a common logarithmic colour scale;
           the residual row uses a diverging scale centred on zero.
           Pixel-to-pixel scatter is $\sim$0.14\,dex for the Balmer lines;
           \OIII\ and \SII\ are systematically elevated.}
  \label{fig:3d_maps}
\end{figure*}

\begin{figure*}[tbp]
  \centering
  \includegraphics[width=\textwidth]{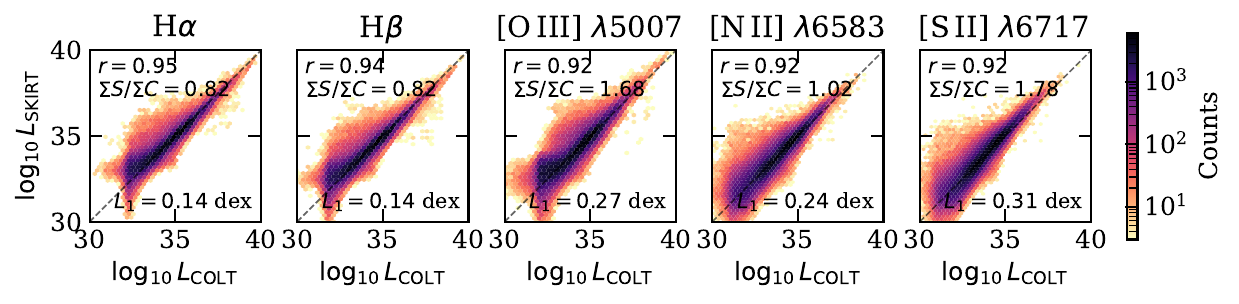}
  \caption{Pixel-by-pixel comparison of \texttt{SKIRT} and \texttt{COLT}
           emission-line luminosities across all four inclinations.
           The dashed line marks perfect agreement.
           The integrated luminosity ratio $\Sigma S / \Sigma C$, Pearson
           correlation coefficient $r$, and luminosity-weighted $L_1$ scatter
           (in dex) are quoted in each panel.
           Pearson $r \geq 0.92$ for all lines; pixel-to-pixel scatter
           is $0.14$--$0.31$\,dex.}
  \label{fig:3d_scatter}
\end{figure*}

Figure~\ref{fig:3d_maps} shows face-on projected maps of H$\alpha$, H$\beta$,
\OIII$\lambda5007$, \NII$\lambda6583$, and \SII$\lambda6717$ from both codes.
The overall morphology is reproduced: both codes identify the same
spiral arm structure, with bright \HII\ region complexes embedded in
extended diffuse emission.
The residual maps (bottom row of Fig.~\ref{fig:3d_maps}) and the
pixel-by-pixel scatter plots (Fig.~\ref{fig:3d_scatter}) quantify the
agreement, with the full statistics listed in Table~\ref{tab:3d_stats}.
The Pearson correlation coefficients are $r \approx 0.94$--$0.95$ for
H$\alpha$ and H$\beta$, and $r \approx 0.92$ for the forbidden lines,
confirming good spatial correspondence across all phases of the ionised ISM.
The Balmer lines agree to within 18 per cent in the integrated luminosity
($\Sigma S / \Sigma C = 0.82$ for both H$\alpha$ and H$\beta$, so the
Balmer decrement is preserved) with the smallest pixel-to-pixel scatter
($L_1 \approx 0.14$\,dex).
\NII\ matches \texttt{COLT} to within 2 per cent in the integrated
luminosity.
The forbidden lines \OIII\ and \SII\ show systematic excesses
($\Sigma S / \Sigma C = 1.68$ and $1.78$, respectively) and
correspondingly larger pixel-to-pixel scatter ($L_1 = 0.27$ and
$0.31$\,dex).
A per-line decomposition of these residuals across the cells where the
two codes overlap and where they do not is presented in
Appendix~\ref{app:perline_decomposition}; in summary, the \OIII\
excess is driven by a modest ($\sim$540\,K) offset in the
emission-weighted gas temperature (\texttt{SKIRT} hotter than
\texttt{COLT}), amplified by the steep temperature dependence of the
collisionally excited line, while the \SII\ excess
is dominated by a population of dense midplane cells in which
\texttt{SKIRT}'s diffuse-LyC treatment likely sustains partial
ionisation that \texttt{COLT}'s on-the-spot Case~B treatment does
not capture.
A residual contribution likely arises in regions where the
underlying hydrodynamical resolution does not fully resolve the
Strömgren-sphere or ionisation-front structure, in which the two
codes' approximations to the local ionisation state can behave
differently.

A further difference between the two models arises in hot, low-density
supernova bubbles: \texttt{COLT} employs Courant-limited cooling
\citep{2025arXiv251013952M} to preserve these structures from the
hydrodynamical simulation, where the low densities suppress
recombination emission and metals are collisionally ionised to
states above those traced by the optical forbidden lines.
\texttt{SKIRT} does not model this non-equilibrium cooling, which
contributes to the residual differences in low-density regions
where shock-heated gas persists in the hydrodynamical input.

\begin{figure*}[tbp]
  \centering
  \includegraphics[width=\textwidth]{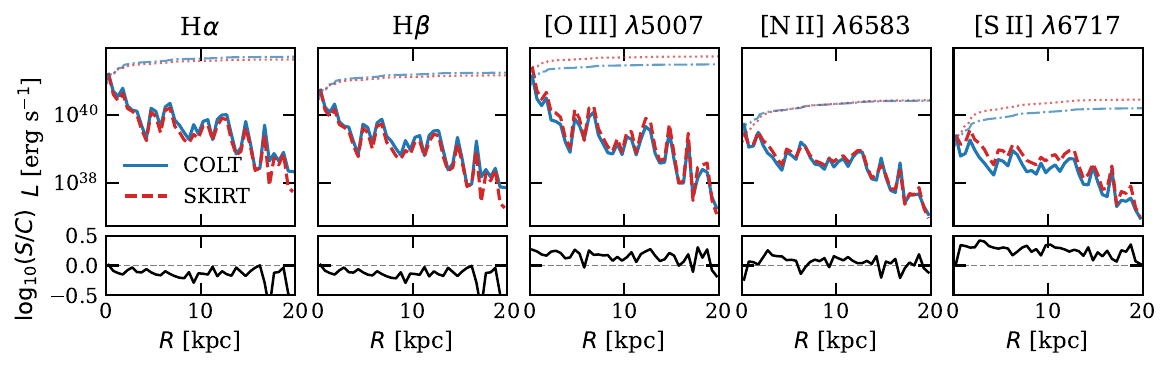}
  \caption{Radial luminosity profiles (face-on, $i = 0^\circ$) for each
           emission line.
           Top sub-panels show the total luminosity per annulus (solid lines)
           and the cumulative luminosity $L(r < R)$ (dash-dot/dotted lines).
           Blue lines are \texttt{COLT}; red lines are \texttt{SKIRT}.
           Bottom sub-panels show the $\log_{10}$ ratio of the annular
           luminosities.
           Both codes place $\sim$80\% of the H$\alpha$ luminosity within
           $R \approx 10$\,kpc.}
  \label{fig:3d_radial}
\end{figure*}

\begin{figure}[tbp]
  \centering
  \includegraphics[width=\columnwidth]{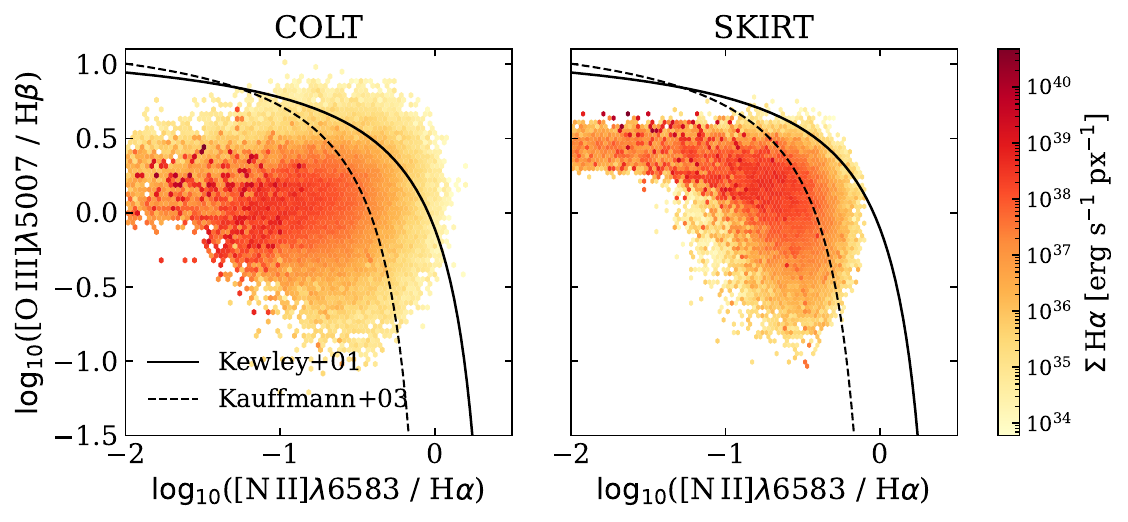}
  \caption{\NII-BPT diagram for the \citet{2020MNRAS.499.5732K}
           Milky Way-analogue galaxy (face-on, $i = 0^\circ$).
           Left: \texttt{COLT}; right: \texttt{SKIRT}.
           Hexbins are coloured by the cumulative H$\alpha$ luminosity.
           The solid and dashed curves show the \citet{2001ApJ...556..121K}
           and \citet{2003MNRAS.346.1055K} demarcation lines, respectively.
           Both codes populate the same star-forming locus of the BPT plane.}
  \label{fig:3d_bpt}
\end{figure}

\begin{table}
  \caption{Summary of statistical agreement between \texttt{SKIRT} and
           \texttt{COLT} for the Milky Way galaxy comparison.
           The ratio $\Sigma S / \Sigma C$ is the integrated luminosity ratio
           summed over all four inclinations; $r$ is the Pearson correlation
           coefficient in log-space; $L_1 = \langle |\log_{10}(S/C)| \rangle_w$ is the mean
           absolute deviation in dex, weighted by the geometric mean
           luminosity $w = \sqrt{S \cdot C}$.}
  \label{tab:3d_stats}
  \centering
  \begin{tabular}{lccc}
    \toprule
    Line & $\Sigma S / \Sigma C$ & Pearson $r$ & $L_1$ [dex] \\
    \midrule
    H$\alpha$\,$\lambda6563$ & $0.82$ & $0.95$ & $0.14$ \\
    H$\beta$\,$\lambda4861$  & $0.82$ & $0.94$ & $0.14$ \\
    \OIII\,$\lambda5007$     & $1.68$ & $0.92$ & $0.27$ \\
    \NII\,$\lambda6583$      & $1.02$ & $0.92$ & $0.24$ \\
    \SII\,$\lambda6717$      & $1.78$ & $0.92$ & $0.31$ \\
    \bottomrule
  \end{tabular}
\end{table}

The spatial distribution of the emission is compared in Fig.~\ref{fig:3d_radial},
which shows the annular luminosity profiles and the cumulative luminosity
enclosed within radius $R$ for all five lines.
The profiles track each other well across the disc for H$\alpha$, H$\beta$,
and \NII.
\OIII\ and \SII\ lie systematically above \texttt{COLT} at all radii;
both offsets are quantified and decomposed in
Appendix~\ref{app:perline_decomposition}.
The cumulative curves confirm that both codes place the bulk of the emission
within the same galactocentric radii: approximately 80 per cent of the total
H$\alpha$ luminosity is enclosed within $R \approx 10$\,kpc in both
\texttt{SKIRT} and \texttt{COLT}.

Figure~\ref{fig:3d_bpt} extends the comparison to the
\citet{1981PASP...93....5B} diagnostic plane, plotting
$\log_{10}(\NII\lambda6583/\mathrm{H}\alpha)$ against
$\log_{10}(\OIII\lambda5007/\mathrm{H}\beta)$ on a pixel-by-pixel basis
for the face-on projection ($i = 0^\circ$), coloured by the cumulative
H$\alpha$ luminosity.
Both \texttt{SKIRT} and \texttt{COLT} populate the star-forming locus of
the Baldwin--Phillips--Terlevich (BPT) plane, with the bulk of the
emission falling below the \citet{2001ApJ...556..121K} demarcation
line, as expected for gas ionised by stellar radiation.
The two distributions overlap across the bulk of this locus but
differ in two systematic ways.
First, \texttt{SKIRT}'s high-luminosity pixels (dominated by \HII\
regions) sit at modestly higher $\log_{10}(\OIII/\mathrm{H}\beta)$;
this is consistent with the cell-by-cell decomposition in
Appendix~\ref{app:perline_decomposition}, which finds \texttt{SKIRT}'s
emission-weighted gas temperature is $\sim$540\,K higher than
\texttt{COLT}'s in cells where both codes emit \OIII, elevating the
line through the steep temperature dependence of the collisionally
excited \OIII.
Second, \texttt{COLT} populates more dim pixels beyond the
\citet{2003MNRAS.346.1055K} and \citet{2001ApJ...556..121K}
demarcations into the composite/LINER regime, and shows a broader
spread along both BPT axes overall.
\texttt{SKIRT}'s table-driven approach pins each cell to its
photoionisation-equilibrium temperature, whereas \texttt{COLT}'s
Courant-limited cooling \citep{2025arXiv251013952M} preserves hot,
low-density gas from the hydrodynamical input.
The resulting wider range of tracked gas temperatures in
\texttt{COLT} translates directly into a broader BPT distribution.
A milder additional effect arises from the diffuse-LyC treatment:
\texttt{SKIRT} explicitly propagates a fraction of the recombination
radiation (Sect.~\ref{sec:reemission}), whereas \texttt{COLT}
assumes on-the-spot (Case~B) reabsorption.
This is expected to shift low-luminosity diffuse pixels toward
higher \OIII/H$\beta$ and lower \NII/H$\alpha$ in \texttt{SKIRT},
but does not dominate the visible distribution.
Marginally-resolved gas in the underlying simulation may also
contribute to the residual differences, since the two codes'
approximations to the local ionisation state behave differently
when the gas structure is not resolved at the cell scale.

%% file: conclusions.tex
\section{Summary and conclusions}
\label{sec:conclusions}

We have presented \texttt{DiffuseIonizedGasMix}, a new material-mix module for
\texttt{SKIRT} that enables photoionisation modelling of diffuse ionised
gas within a full three-dimensional MCRT framework, using a hybrid approach
that combines pre-computed \texttt{Cloudy} tables with an inline solver.
The module characterises the local radiation field using the ionisation parameter
$\log U$ and four spectral-shape ratios across the ionising continuum.
Pre-computed \texttt{Cloudy} tables map this representation to gas temperature
and absorption opacities, while emission-line luminosities are computed
from ion fractions determined by an inline solver at the converged temperature.
The ionisation state converges with the radiation field through the
existing \texttt{SKIRT} iteration cycle, with no modifications to the MCRT engine.

We validated the module in two stages.
In the first stage (Sect.~\ref{sec:results_1d}), we compared integrated line luminosities
against \texttt{Cloudy} for a grid of 60 one-dimensional spherical shell models spanning
five ionising photon rates ($Q_{\rm ion} = 10^{48}$--$10^{50}$\,s$^{-1}$), three
hydrogen number densities ($n_{\rm H} = 1$--$100$\,cm$^{-3}$), and four stellar
population ages (1\,Myr--1\,Gyr) at solar metallicity.
The hydrogen recombination lines agree with \texttt{Cloudy} to within a few per cent
in the median (H$\alpha$ median 0.97, H$\beta$ median 0.94), and the
forbidden lines \OIII\ and \NII\ show median ratios of 1.02 and 1.04,
respectively.
\SII\ has a median ratio of 1.23, the largest offset, driven by the
temperature overestimate near the ionisation front where this line is produced.
We also compared against \texttt{COLT} \citep{2025arXiv251013952M} run on a
one-dimensional spherical grid with shell edges matched to the \texttt{Cloudy} zone boundaries.
Both codes reproduce the metal lines to within $\sim$10\% of \texttt{Cloudy}
in the median and recover the Balmer lines at comparable levels.

In the second stage (Sect.~\ref{sec:results_3d}), we applied \texttt{DiffuseIonizedGasMix}
to a hydrodynamical simulation of an isolated Milky Way-analogue galaxy
\citep{2020MNRAS.499.5732K} and compared the resulting emission-line maps against the
\texttt{COLT} run on the full three-dimensional Voronoi mesh.
The pixel-by-pixel Pearson correlation coefficients are $r \approx 0.94$--$0.95$
for H$\alpha$ and H$\beta$, and $r \approx 0.92$ for the forbidden lines
\OIII, \NII, and \SII.
The integrated luminosity ratios are 0.82 for the hydrogen recombination
lines (so the Balmer decrement is preserved) and essentially unity for
\NII.
The collisionally excited forbidden lines \OIII\ and \SII\ are
systematically elevated ($\Sigma S/\Sigma C = 1.68$ and $1.78$,
respectively); the per-line decomposition presented in
Appendix~\ref{app:perline_decomposition} traces these excesses to a
modest emission-weighted temperature offset (amplified by the steep
temperature dependence of the collisionally excited \OIII\ line) and
to a population of dense midplane cells in which \texttt{SKIRT}'s
diffuse-LyC treatment likely sustains partial ionisation that
\texttt{COLT}'s on-the-spot Case~B treatment does not capture.
We also note that some of the residual scatter likely reflects the
finite spatial resolution of the underlying simulation; higher-resolution
input would help separate genuine modelling differences from those tied
to unresolved \HII\ regions and ionisation fronts.
Refinement strategies guided by the local gas state will be explored in
a forthcoming paper of this series (Kapoor et al., in prep.).

These tests demonstrate that \texttt{DiffuseIonizedGasMix} provides a physically
consistent route to predicting synthetic emission-line
observables of ionised gas in \texttt{SKIRT}, kept computationally tractable
by the use of pre-computed tables for the temperature and opacity convergence.
As emission lines, dust attenuation, and dust re-emission are computed in a single MCRT run, the module provides a complete forward-modelling pipeline for arbitrary three-dimensional geometries that can be applied directly to the output of any hydrodynamical simulation via \texttt{SKIRT}'s native snapshot import.

\subsection*{Limitations and future directions}
\label{sec:outlook}

The average gas temperature recovered by the five-bin radiation field STAB tables agrees with
\texttt{Cloudy} to within $\sim$3\% across the benchmark grid,
demonstrating that the coarse spectral characterisation captures the dominant
heating and cooling processes well.
However, the tables currently assume solar-scaled abundance ratios, so the
temperature and opacity are fixed to a single set of relative element
proportions regardless of the actual composition of each gas cell.
Solving the full heating--cooling balance from the wavelength-resolved
radiation field would lift this fixed-abundance constraint and improve the
temperature accuracy near ionisation fronts, where the residual is largest,
at the cost of additional computation and generally more iterations.

A second class of physics not currently included in the tables is the
contribution from dust grains.
The STAB tables are computed in the absence of dust, but this does not
preclude including dust within \texttt{SKIRT}'s photon cycle.
\texttt{SKIRT} is natively a dust radiative-transfer code, and the gas
medium can be co-deployed with dust media: the photon cycle then transports
radiation through both, and each gas cell sees a locally dust-attenuated
radiation field.
The same dust-free STAB tables can be queried with this dust-modified
$\log U$ and spectral-shape ratios, capturing the leading effect of dust on
the gas ionisation and temperature without any modification to the tables
themselves.
A more thorough treatment, extending the thermal balance to include direct
dust-driven channels such as photoelectric heating from grain surfaces
\citep{1994ApJ...427..822B, 2001ApJS..134..263W} and gas-grain collisional
coupling \citep{1983ApJ...265..223B}, is left for future work.

Modern cosmological simulations routinely track per-cell abundances for
these elements (e.g.\ \texttt{EAGLE}, \citealt{2015MNRAS.446..521S};
\texttt{IllustrisTNG}, \citealt{2018MNRAS.473.4077P};
\texttt{COLIBRE}, \citealt{2025arXiv250821126S}).
Since the inline solver already computes ion fractions from individual elemental
abundances, the module reads per-cell abundances directly from the simulation
snapshot; the main challenge lies instead in the temperature and opacity tables,
which currently
assume solar-scaled ratios: adding individual element abundances as free
parameters would increase the table dimensionality considerably.
However, certain non-solar patterns that can be parameterised with a small number
of additional dimensions (e.g.\ an $\alpha$-enhancement or N/O ratio) could be
incorporated without a prohibitive increase in table size, because these
patterns add only one or two dimensions rather than the full set of
individual element abundances; this is the route taken in the next paper of
this series (Kapoor et al., in prep.).
A complementary approach is explored by Lauwers et al.\ (in prep.), who
execute \texttt{Cloudy} simulations on demand during the radiative transfer
calculation: whenever \texttt{SKIRT} encounters a combination of metallicity,
density, and radiation-field parameters not yet present in the table, a
\texttt{Cloudy} run is launched to fill the gap.
The resulting outputs progressively populate a high-dimensional table,
concentrating computational effort on the regions of parameter space that are
actually sampled and enabling a finer representation of the local conditions
without requiring a densely pre-computed grid.

\subsection*{Potential applications}
\label{sec:applications}

\texttt{DiffuseIonizedGasMix} opens several avenues
that were previously inaccessible within \texttt{SKIRT}.

\textit{Mock IFU observations.}
The module produces emission-line maps in the same run that computes
dust attenuation and thermal emission, enabling direct mock-IFU comparisons
(e.g.\ MUSE, \texttt{JWST} NIRSpec/IFU) without post-processing corrections
for ionised-gas emission.
The MCRT framework naturally accounts for geometry, inclination effects,
and the high dynamic range and multi-phase structure of the ISM, so this forward-modelling approach avoids
the systematic biases inherent in inverse methods that derive physical quantities
from observations.

\textit{BPT diagnostics across cosmic time.}
\texttt{JWST} has revealed that high-redshift galaxies are systematically offset
from the local BPT locus toward higher \OIII/H$\beta$ at fixed
\NII/H$\alpha$ \citep{2025ApJ...980..242S}.
Several explanations have been proposed, including non-solar abundance patterns,
harder ionising spectra, and diffuse ionised gas emission, which contributes up
to $\sim$60\% of total H$\alpha$ luminosity in nearby star-forming galaxies
\citep{2009RvMP...81..969H} but whose role at high redshift is essentially unconstrained.
By post-processing cosmological simulations at multiple redshifts and systematically
varying the ionising spectrum (e.g.\ binary populations, stripped-star models,
AGN contributions), the module enables controlled experiments that isolate the
contributions of the ionising SED, gas-phase abundances, and 3D geometry to the
observed BPT evolution.

\textit{Multi-wavelength SEDs.}
The unified treatment of stars, gas, and dust in a single simulation enables the
construction of physically consistent spectral energy distributions from the
ionising UV through the far-infrared for populations of simulated galaxies,
making the module directly applicable to forward-modelling pipelines for large
cosmological simulations and upcoming wide-field IFU surveys.

These methodological extensions and applications form a programme of
forthcoming papers in this series.

%% file: appendix.tex
\begin{appendix}

\section{Convergence of the Milky Way simulation}
\label{app:convergence}

\begin{figure}
  \centering
  \includegraphics[width=\columnwidth]{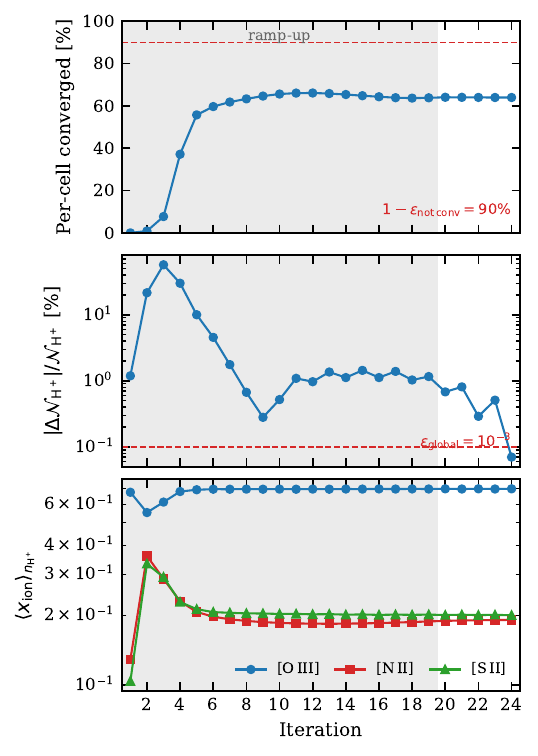}
  \caption{Convergence of the Milky Way-analogue simulation
           (Sect.~\ref{sec:results_3d}).
           \textit{Top:} fraction of cells satisfying the per-cell
           convergence criterion (relative change in $T$ and $\log U$
           below $\epsilon_{\rm cell} = 0.01$) as a function of iteration.
           \textit{Middle:} relative change in total ionised hydrogen
           ($\Delta \mathcal{N}_{\mathrm{H}^+} / \mathcal{N}_{\mathrm{H}^+}$)
           per iteration on a logarithmic scale.
           \textit{Bottom:} $n_{\mathrm{H}^+}$-weighted mean ion fractions
           of \OIII, \NII, and \SII\ across iterations.
           Dashed lines in the top two panels mark the convergence
           thresholds from Table~\ref{tab:mw_config}; the shaded region
           indicates the photon-packet ramp-up phase.
           Convergence is reached at iteration~24.}
  \label{fig:convergence}
\end{figure}

Figure~\ref{fig:convergence} shows the convergence behaviour of the
\texttt{DiffuseIonizedGasMix} iteration cycle for the Milky Way-analogue
simulation described in Sect.~\ref{sec:results_3d_setup}.
Since the per-cell state update (a table interpolation) is fast
compared to the Monte Carlo photon propagation, the total runtime
scales primarily with the number of photon packets times the number of
iterations.
Early iterations serve mainly to establish the broad spatial structure of the
radiation field, so fewer packets suffice.
The run therefore uses photon-packet ramp-up: the user specifies an initial
packet fraction and a target iteration at which the full budget is reached;
the number of packets then increases geometrically between these two
endpoints.
Here the count grows from $10^7$ in the first iteration to the full $10^9$
by iteration~19 (shaded region in Fig.~\ref{fig:convergence}).
Each early iteration uses only a fraction of the full packet
budget, so the 18 ramped iterations together cost roughly the same as four iterations at full packet count, while providing 18 convergence steps.

The per-cell converged fraction (top panel) rises during the first few
iterations and then plateaus near $\sim$64\%.
The bulk of the unconverged cells are low-ionisation
($\log U \lesssim -3$) Voronoi cells in which the small absolute
photoionisation rate is dominated by Monte Carlo noise: ($T$, $\log U$)
bounce at amplitudes just above the stringent per-cell threshold
($\epsilon_{\rm cell} = 0.01$) without affecting the integrated emission,
so the run effectively converges through the global criterion rather
than through the per-cell criterion.
The global ionised-hydrogen change (middle panel) is non-monotonic
during the photon-packet ramp-up: it falls to a local minimum of
$\sim 0.3\%$ at iteration~9 and then rises into the
$\sim 1$--$1.5\%$ range as the photon count continues to climb.
This reflects the rising budget: each successive ramp iteration delivers
a more accurate, less noise-limited update that can genuinely perturb
the gas state, so a stricter convergence criterion is temporarily harder
to satisfy.
Stopping at the apparent dip near iteration~9 would therefore be
premature; once the ramp ends at iteration~19 the criterion settles
and drops below the $10^{-3}$ threshold by iteration~24, at which point
convergence is declared.
The bottom panel shows the $n_{\mathrm{H}^+}$-weighted mean ion fractions
of three diagnostic species (\OIII, \NII, \SII) tracked across the
iteration cycle; their values flatten well before the global criterion
is met, confirming that the emission-line predictions are stable at
the converged solution.

\section{Bin-count sensitivity test}
\label{app:bin_sensitivity}

To quantify the temperature and photon-absorption-rate errors arising from
the five-bin radiation-field characterisation, we reconstruct the input SED
at progressively finer bin counts and compare single-zone \texttt{Cloudy}
outputs.
For each (age, $\log U$, $n_{\rm H}$) configuration on a 72-point grid (4
stellar ages from 1\,Myr to 1\,Gyr, 6 ionisation parameters
$\log U \in [-3.0,\,-0.5]$, 3 hydrogen densities
$n_{\rm H} \in \{1,\,10,\,100\}\,\mathrm{cm}^{-3}$) at $Z = 0.014$ with
GASS10 abundances, we run \texttt{Cloudy} once with the original
\texttt{BPASS} chab100 SED and again with step-function reconstructions at
2, 3, 5, 10, and 20 bins.
The bin layouts at coarser counts are obtained by progressively dropping
internal boundaries from the paper's 5-bin scheme
(1.00, 1.80, 2.58, 3.52, 4.00, 6.00\,Ryd) so that each bin transition
corresponds to a physical ionisation edge; the higher counts subdivide each
paper bin uniformly.
For each pair of runs we compute the relative error in the equilibrium
temperature $T_{\rm e}$ and in the photon-absorption rate
$\int L_{\rm full}(\nu)\,\kappa(\nu)\,\mathrm{d}\nu$ over $[1,\,6]$\,Ryd, where
$\kappa(\nu)$ is the gas absorption opacity at \texttt{Cloudy}'s native
$R \sim 300$ spectral resolution and $L_{\rm full}(\nu)$ is the unbinned
\texttt{BPASS} spectrum, so that the bin compression enters only through the
gas state.

\begin{table}
  \caption{Bin-count sensitivity: relative errors between the original
           \texttt{BPASS} SED and step-function reconstructions, in the
           equilibrium temperature ($T_{\rm e}$) and in the
           photon-absorption rate
           $\int L_{\rm full}(\nu)\,\kappa(\nu)\,\mathrm{d}\nu$ over $[1,\,6]$\,Ryd,
           averaged over 72 single-zone configurations.}
  \label{tab:bin_sensitivity}
  \centering
  \begin{tabular}{ccc}
    \toprule
    $N_{\rm bins}$ & $\Delta T_{\rm e}$ (\%) &
    $\Delta\!\int\!L_{\rm full}\kappa\,\mathrm{d}\nu$ (\%) \\
    \midrule
    2  & 7.8 & 20.9 \\
    3  & 6.6 & 15.1 \\
    5  & 6.2 &  5.9 \\
    10 & 4.5 &  3.6 \\
    20 & 3.8 &  3.2 \\
    \bottomrule
  \end{tabular}
\end{table}

\begin{figure}
  \centering
  \includegraphics[width=\columnwidth]{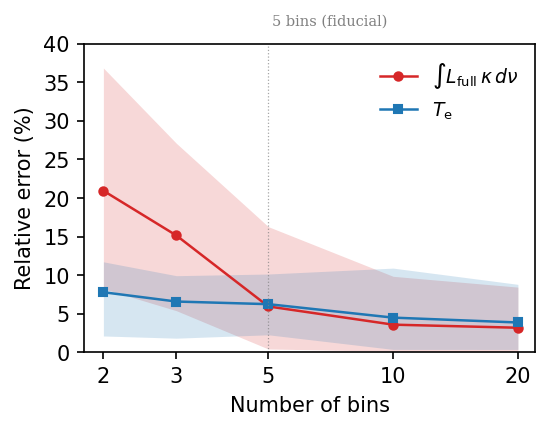}
  \caption{Bin-count sensitivity test: mean relative error in equilibrium
           temperature ($T_{\rm e}$, blue squares) and in the
           photon-absorption rate
           $\int L_{\rm full}(\nu)\,\kappa(\nu)\,\mathrm{d}\nu$ (red circles) over
           $[1,\,6]$\,Ryd between the original \texttt{BPASS} SED and
           step-function reconstructions at progressively finer bin counts.
           Shaded bands show the $16$--$84$ percentile spread across the
           72 single-zone configurations.
           The dotted vertical line marks the paper's fiducial 5-bin
           choice.}
  \label{fig:bin_sensitivity}
\end{figure}

The 5-bin scheme adopted in this paper sits at the elbow of the
photon-absorption-rate curve, where the relative error drops sharply from
$15\%$ at 3 bins to $6\%$ at 5 and only marginally further at higher counts;
the temperature error declines more gradually across the entire range.
The modest accuracy gains from doubling the bin count to 10 do not justify
the resulting STAB table-size explosion (a factor of $\sim 6 \times 10^4$
at the current node density).
Individual line ratios sensitive to the spectral shape (notably the
temperature-sensitive forbidden lines such as \SII) could in principle
benefit further from a finer binning; the 1D \texttt{Cloudy} benchmark
in Sect.~\ref{sec:results_1d} shows that at 5 bins the per-line
luminosity errors are already at the level expected from the 1D
\texttt{Cloudy} agreement.

\section{Performance comparison}
\label{app:speed}

We complement the qualitative discussion in
Sect.~\ref{sec:DIG_role} with a quantitative wall-time measurement on
a 1D Strömgren-sphere test configured identically for \texttt{SKIRT}
and \texttt{COLT}.
The chosen model is
$\mathrm{Q}_{\mathrm{ion}} = 10^{49}\,\mathrm{s}^{-1}$,
$n_{\rm H} = 10\,\mathrm{cm}^{-3}$, $Z = 0.014$ with GASS10
abundances, and a 10\,Myr \texttt{BPASS} chab100 SED.
Both codes propagate radiation through 1250 linearly-spaced radial
shells, starting from a fully-ionised warm initial condition, on the
same machine using 8 OpenMP threads, with the convergence thresholds
adopted for the production Milky Way run (Table~\ref{tab:mw_config}).
\texttt{SKIRT} starts the convergence loop at 1\% of the full photon
budget and grows the count geometrically by a factor of $1.3$ per
iteration; the \texttt{COLT} run is configured with five pre-conditioning
iterations before its main solve.
A direct comparison on the 3D Milky Way run of
Sect.~\ref{sec:results_3d} is complicated by the two codes employing
different photon budgets, line treatments, and output formats,
motivating this 1D test as the cleanest setting in which both can be
configured on identical input.

Table~\ref{tab:speed_1d} reports the wall-time breakdown of the
apples-to-apples portion: \texttt{SKIRT} reaches the converged gas
state in 688\,s, $\sim$1.5 times faster than \texttt{COLT} (1027\,s).
The line-emission stages that follow in each code are not included in
this comparison: \texttt{SKIRT} produces all 20 lines and the full SED
in a single secondary-emission iteration, whereas \texttt{COLT} runs a
separate spectrally-resolved MCRT for each requested line, so the
two stages differ in scope and are not directly comparable.
The \texttt{SKIRT} convergence loop itself accounts for only 115\,s
(12 ramped iterations); the rest of the primary phase is the final
full-budget emission iteration that establishes the radiation field
at the photon count required for the downstream line stage.
\texttt{COLT} runs 17 iterations in total at fixed per-iteration
photon count (five pre-conditioning iterations followed by 12 main-solve
iterations), so its wall-time scales linearly with the iteration count.
The per-cell physics step (table interpolation for \texttt{SKIRT},
thermal/ionisation solution for \texttt{COLT}) is small compared to
photon propagation in both codes; the wall-time difference reflects
\texttt{SKIRT}'s faster convergence, consistent with the discussion
in Sect.~\ref{sec:DIG_role}.
We note that the \texttt{COLT} ramp-up was not specifically tuned for
this comparison.

\begin{table}
  \caption{Wall-time breakdown on the 1D Strömgren-sphere test
           ($\mathrm{Q}_{\mathrm{ion}} = 10^{49}\,\mathrm{s}^{-1}$,
           $n_{\rm H} = 10\,\mathrm{cm}^{-3}$, $Z = 0.014$, 10\,Myr
           \texttt{BPASS} chab100 SED, 1250 shells, 8 threads).
           Listed phases produce the converged gas state in each code
           and are directly comparable; the subsequent line-emission
           stages differ in scope between the two codes and are not
           included.}
  \label{tab:speed_1d}
  \centering
  \begin{tabular}{@{}lr@{}}
    \toprule
    Phase & Wall time \\
    \midrule
    \texttt{SKIRT} convergence loop (12 ramped iterations) & 115\,s \\
    \texttt{SKIRT} final primary emission iteration        & 573\,s \\
    \texttt{SKIRT} primary phase total                     & \textbf{688\,s} \\[2pt]
    \texttt{COLT} pre-conditioning + ionisation solve      & \textbf{1027\,s} \\
    \bottomrule
  \end{tabular}
\end{table}

\section{Per-line decomposition of the 3D Milky Way comparison}
\label{app:perline_decomposition}

To trace the systematic \OIII\ and \SII\ excesses reported in
Sect.~\ref{sec:results_3d_gal} to specific gas populations, we
performed a cell-by-cell comparison on the matched Voronoi mesh.
Two complementary masks are used: the \emph{overlap} mask, containing
cells where both codes emit a given line, and the
\emph{\texttt{COLT}-non-emitting} mask, containing cells where
\texttt{SKIRT} emits but the matched \texttt{COLT} cell does not.
The overlap mask captures $\geq 90\%$ of \texttt{COLT}'s integrated
luminosity for every line; for \texttt{SKIRT} it captures
$\geq 87\%$ for the recombination lines and \OIII, but only $54\%$
for \SII\ and $63\%$ for \NII, so the \SII\ and \NII\ stories
require both masks.
Roughly $7.3 \times 10^5$ cells satisfy the \texttt{COLT}-non-emitting
condition; $55\%$ of them are pinned at the $300$\,K temperature
floor adopted in this \texttt{COLT} run.

For \OIII$\lambda5007$, the overlap-mask ion-fraction ratio is
unity to within $6\%$, but the emission-weighted temperature is
$\sim 540$\,K higher in \texttt{SKIRT} ($7469$\,K vs $6928$\,K).
Since \OIII\ is collisionally excited and emits with a steep
dependence on gas temperature, this offset shifts the line emission
upward in \texttt{SKIRT} and accounts for most of the observed
$L_{\rm S}/L_{\rm C} = 1.50$ overlap excess.
The overlap mask captures $\geq 87\%$ of the \OIII\ emission in
both codes, so the overlap analysis is essentially the whole \OIII\
story.

For \SII$\lambda6717$, the overlap-mask luminosity ratio is balanced
($1.03$), so the full $1.78$ global excess originates in the
\texttt{COLT}-non-emitting cells.
The dominant luminosity carriers within this set are dense midplane
cells with $n_{\rm H} > 1\,\mathrm{cm}^{-3}$
($4.4 \times 10^4$ cells of $\sim$17\,pc cell size, with
$\langle |z| \rangle = 0.18$\,kpc and $\langle R \rangle = 6$\,kpc),
which alone contribute $39\%$ of \texttt{SKIRT}'s total \SII\
luminosity.
In those cells the ion fraction $x_{\rm S\,II}$ is comparable in the
two codes, but \texttt{COLT} does not emit because the cells are
pinned at $300$\,K (Boltzmann suppression of the collisional
excitation),
whereas \texttt{SKIRT} returns a warm partially-ionised state
($T \sim 6600$\,K, $x_{\rm HII} \sim 0.55$ emission-weighted).
We attribute the difference to the diffuse-LyC treatment:
\texttt{SKIRT} re-emits absorbed LyC photons
(Sect.~\ref{sec:reemission}), heating and partially ionising the
shielded midplane gas to conditions warm enough to sustain
collisional excitation of \SII, whereas \texttt{COLT}'s on-the-spot
Case~B treatment does not transport this diffuse field, and the
same cells cool to the temperature floor and stop emitting.

For \NII$\lambda6583$, \texttt{SKIRT}'s overlap-mask luminosity is
$32\%$ lower than \texttt{COLT}'s, driven by a lower ion-fraction
ratio ($\langle x_{\rm S}/x_{\rm C} \rangle = 0.42$); the gas
temperatures agree to within $1\%$, so the difference is not
temperature-driven and reflects how the two codes partition nitrogen
across ionisation stages at intermediate $\log U$.
The dense \texttt{COLT}-non-emitting midplane cells discussed above
contribute $\sim 31\%$ of \texttt{SKIRT}'s total \NII\ luminosity
through the same diffuse-LyC mechanism, raising the integrated
ratio to $\Sigma S/\Sigma C = 1.02$.
The integrated agreement therefore combines an overlap-mask
ion-fraction deficit and a comparable \texttt{COLT}-non-emitting boost,
two physically distinct effects of similar magnitude.

\begin{table}
  \caption{Overlap-mask statistics for the cell-by-cell \texttt{SKIRT}/\texttt{COLT}
           comparison of the \OIII, \SII, and \NII\ ratios.
           $N_{\rm ov}$: number of overlap cells.
           $\%L_{\rm S}$, $\%L_{\rm C}$: fraction of the integrated
           \texttt{SKIRT}/\texttt{COLT} luminosity captured by the
           overlap mask.
           $L_{\rm S}/L_{\rm C}$: overlap luminosity ratio.
           $\langle x_{\rm S}/x_{\rm C}\rangle$: emission-weighted
           mean \texttt{SKIRT}-to-\texttt{COLT} ion-fraction ratio.
           $\langle T_{\rm S}\rangle / \langle T_{\rm C}\rangle$:
           emission-weighted mean cell temperatures.}
  \label{tab:perline_overlap}
  \centering
  \small
  \setlength{\tabcolsep}{4pt}
  \begin{tabular}{@{}lcccccc@{}}
    \toprule
    Line & $N_{\rm ov}$ & $\%L_{\rm S}$ & $\%L_{\rm C}$ &
    $L_{\rm S}/L_{\rm C}$ & $\langle x_{\rm S}/x_{\rm C}\rangle$ &
    $\langle T_{\rm S}\rangle/\langle T_{\rm C}\rangle$ [K] \\
    \midrule
    \OIII\,$\lambda5007$ & 0.86M & 87 & 97 & 1.50 & 1.06 & 7469 / 6928 \\
    \SII\,$\lambda6716$  & 0.93M & 54 & 93 & 1.03 & 0.89 & 7343 / 7109 \\
    \NII\,$\lambda6583$  & 0.86M & 63 & 94 & 0.68 & 0.42 & 7408 / 7338 \\
    \bottomrule
  \end{tabular}
\end{table}

\end{appendix}

%% file: bibliography.bib
@preamble{ " \newcommand{\noop}[1]{} " }

@ARTICLE{2024A&A...691A..19V,
       author = {{Venturi}, G. and {Carniani}, S. and {Parlanti}, E. and {Kohandel}, M. and {Curti}, M. and {Pallottini}, A. and {Vallini}, L. and {Arribas}, S. and {Bunker}, A.~J. and {Cameron}, A.~J. and {Castellano}, M. and {Ferrara}, A. and {Fontana}, A. and {Gallerani}, S. and {Gelli}, V. and {Maiolino}, R. and {Ntormousi}, E. and {Pacifici}, C. and {Pentericci}, L. and {Salvadori}, S. and {Vanzella}, E.},
        title = "{Gas-phase metallicity gradients in galaxies at z {\ensuremath{\sim}} 6{\textendash}8}",
      journal = {\aap},
         year = 2024,
       volume = {691},
          eid = {A19},
        pages = {A19},
          doi = {10.1051/0004-6361/202449855},
}

@ARTICLE{2025arXiv251016116F,
       author = {{Fujimoto}, Seiji and {Faisst}, Andreas L. and {Tsujita}, Akiyoshi and {Kohandel}, Mahsa and {Lee}, Lilian L. and {{\"U}bler}, Hannah and {Loiacono}, Federica and {Nezhad}, Negin and {Pallottini}, Andrea and {Aravena}, Manuel and {Assef}, Roberto J. and {Battisti}, Andrew J. and {B{\'e}thermin}, Matthieu and {Boquien}, M{\'e}d{\'e}ric and {da Cunha}, Elisabete and {Ferrara}, Andrea and {Franco}, Maximilien and {Ginolfi}, Michele and {Hadi}, Ali and {Haghjoo}, Aryana and {Herrera-Camus}, Rodrigo and {Inami}, Hanae and {Koekemoer}, Anton M. and {Lemaux}, Brian C. and {Li}, Yuan and {Liu}, Lun-Jun and {Molina}, Juan and {Nanni}, Ambra and {Pozzi}, Francesca and {Relano}, Monica and {Romano}, Michael and {Sanders}, David B. and {F{\"o}rster Schreiber}, Natascha M. and {Silverman}, John and {Spilker}, Justin and {Telikova}, Kseniia and {Villanueva}, Vicente and {Vallini}, Livia and {Wang}, Wuji and {Zamorani}, Giovanni},
        title = "{The ALPINE-CRISTAL-JWST Survey: NIRSpec IFU Data Processing and Spatially-resolved Views of Chemical Enrichment in Normal Galaxies at z=4-6}",
      journal = {arXiv e-prints},
         year = 2025,
          eid = {arXiv:2510.16116},
        pages = {arXiv:2510.16116},
          doi = {10.48550/arXiv.2510.16116},
}

@article{2023A&A...678A.175M,
  author        = {{Matsumoto}, Kosei and {Camps}, Peter and {Baes}, Maarten and {De Ceuster}, Frederik and {Wada}, Keiichi and {Nakagawa}, Takao and {Nagamine}, Kentaro},
  title         = {{Self-consistent dust and non-LTE line radiative transfer with SKIRT}},
  journal       = {\aap},
  year          = 2023,
  month         = oct,
  volume        = {678},
  eid           = {A175},
  pages         = {A175},
  doi           = {10.1051/0004-6361/202347376},
  archiveprefix = {arXiv},
  eprint        = {2309.02628},
  primaryclass  = {astro-ph.GA},
  adsurl        = {https://ui.adsabs.harvard.edu/abs/2023A&A...678A.175M}
}

@article{2023A&A...674A.123V,
  author        = {{Vander Meulen}, Bert and {Camps}, Peter and {Stalevski}, Marko and {Baes}, Maarten},
  title         = {{X-ray radiative transfer in full 3D with SKIRT}},
  journal       = {\aap},
  year          = 2023,
  month         = jun,
  volume        = {674},
  eid           = {A123},
  pages         = {A123},
  doi           = {10.1051/0004-6361/202245783},
  archiveprefix = {arXiv},
  eprint        = {2304.10563},
  primaryclass  = {astro-ph.HE},
  adsurl        = {https://ui.adsabs.harvard.edu/abs/2023A&A...674A.123V}
}

@article{2017A&A...601A..92P,
  author        = {{Peest}, C. and {Camps}, P. and {Stalevski}, M. and {Baes}, M. and {Siebenmorgen}, R.},
  title         = {{Polarization in Monte Carlo radiative transfer and dust scattering polarization signatures of spiral galaxies}},
  journal       = {\aap},
  year          = 2017,
  month         = may,
  volume        = {601},
  eid           = {A92},
  pages         = {A92},
  doi           = {10.1051/0004-6361/201630157},
  archiveprefix = {arXiv},
  eprint        = {1702.07354},
  primaryclass  = {astro-ph.IM},
  adsurl        = {https://ui.adsabs.harvard.edu/abs/2017A&A...601A..92P}
}

@article{2024A&A...689A..13L,
  author        = {{Lauwers}, Arno and {Baes}, Maarten and {Camps}, Peter and {Vander Meulen}, Bert},
  title         = {{Tetrahedral grids in Monte Carlo radiative transfer}},
  journal       = {\aap},
  year          = 2024,
  month         = sep,
  volume        = {689},
  eid           = {A13},
  pages         = {A13},
  doi           = {10.1051/0004-6361/202450658},
  archiveprefix = {arXiv},
  eprint        = {2407.20216},
  primaryclass  = {astro-ph.GA},
  adsurl        = {https://ui.adsabs.harvard.edu/abs/2024A&A...689A..13L}
}

@book{2006agna.book.....O,
  author    = {{Osterbrock}, Donald E. and {Ferland}, Gary J.},
  title     = {{Astrophysics of gaseous nebulae and active galactic nuclei}},
  year      = 2006,
  publisher = {{University Science Books}},
  adsurl    = {https://ui.adsabs.harvard.edu/abs/2006agna.book.....O}
}

@article{2018MNRAS.473.4077P,
  author        = {{Pillepich}, Annalisa and {Springel}, Volker and {Nelson}, Dylan and {Genel}, Shy and {Naiman}, Jill and {Pakmor}, R{\"u}diger and {Hernquist}, Lars and {Torrey}, Paul and {Vogelsberger}, Mark and {Weinberger}, Rainer and {Marinacci}, Federico},
  title         = {{Simulating galaxy formation with the IllustrisTNG model}},
  journal       = {\mnras},
  year          = 2018,
  month         = jan,
  volume        = {473},
  number        = {3},
  pages         = {4077-4106},
  doi           = {10.1093/mnras/stx2656},
  archiveprefix = {arXiv},
  eprint        = {1703.02970},
  primaryclass  = {astro-ph.GA},
  adsurl        = {https://ui.adsabs.harvard.edu/abs/2018MNRAS.473.4077P}
}

@article{2015MNRAS.446..521S,
  author        = {{Schaye}, Joop and {Crain}, Robert A. and {Bower}, Richard G. and {Furlong}, Michelle and {Schaller}, Matthieu and {Theuns}, Tom and {Dalla Vecchia}, Claudio and {Frenk}, Carlos S. and {McCarthy}, I.~G. and {Helly}, John C. and {Jenkins}, Adrian and {Rosas-Guevara}, Y.~M. and {White}, Simon D.~M. and {Baes}, Maarten and {Booth}, C.~M. and {Camps}, Peter and {Navarro}, Julio F. and {Qu}, Yan and {Rahmati}, Alireza and {Sawala}, Till and {Thomas}, Peter A. and {Trayford}, James},
  title         = {{The EAGLE project: simulating the evolution and assembly of galaxies and their environments}},
  journal       = {\mnras},
  year          = 2015,
  month         = jan,
  volume        = {446},
  number        = {1},
  pages         = {521-554},
  doi           = {10.1093/mnras/stu2058},
  archiveprefix = {arXiv},
  eprint        = {1407.7040},
  primaryclass  = {astro-ph.GA},
  adsurl        = {https://ui.adsabs.harvard.edu/abs/2015MNRAS.446..521S}
}

@article{2003PASP..115..763C,
  author        = {{Chabrier}, Gilles},
  title         = {{Galactic Stellar and Substellar Initial Mass Function}},
  journal       = {\pasp},
  year          = 2003,
  month         = jul,
  volume        = {115},
  number        = {809},
  pages         = {763-795},
  doi           = {10.1086/376392},
  archiveprefix = {arXiv},
  eprint        = {astro-ph/0304382},
  primaryclass  = {astro-ph},
  adsurl        = {https://ui.adsabs.harvard.edu/abs/2003PASP..115..763C}
}

@article{2017RMxAA..53..385F,
  author        = {{Ferland}, G.~J. and {Chatzikos}, M. and {Guzm{\'a}n}, F. and {Lykins}, M.~L. and {van Hoof}, P.~A.~M. and {Williams}, R.~J.~R. and {Abel}, N.~P. and {Badnell}, N.~R. and {Keenan}, F.~P. and {Porter}, R.~L. and {Stancil}, P.~C.},
  title         = {{The 2017 Release Cloudy}},
  journal       = {\rmxaa},
  year          = 2017,
  month         = oct,
  volume        = {53},
  pages         = {385-438},
  archiveprefix = {arXiv},
  eprint        = {1705.10877},
  primaryclass  = {astro-ph.GA},
  adsurl        = {https://ui.adsabs.harvard.edu/abs/2017RMxAA..53..385F}
}

@article{2010Ap&SS.328..179G,
  author   = {{Grevesse}, N. and {Asplund}, M. and {Sauval}, A.~J. and {Scott}, P.},
  title    = {{The chemical composition of the Sun}},
  journal  = {\apss},
  year     = 2010,
  month    = jul,
  volume   = {328},
  number   = {1-2},
  pages    = {179-183},
  doi      = {10.1007/s10509-010-0288-z},
  adsurl   = {https://ui.adsabs.harvard.edu/abs/2010Ap&SS.328..179G}
}

@article{1981PASP...93....5B,
  author   = {{Baldwin}, J.~A. and {Phillips}, M.~M. and {Terlevich}, R.},
  title    = {{Classification parameters for the emission-line spectra of extragalactic objects.}},
  journal  = {\pasp},
  year     = 1981,
  month    = feb,
  volume   = {93},
  pages    = {5-19},
  doi      = {10.1086/130766},
  adsurl   = {https://ui.adsabs.harvard.edu/abs/1981PASP...93....5B}
}

@article{2016A&A...590A..55B,
  author        = {{Baes}, M. and {Gordon}, K.~D. and {Lunttila}, T. and {Bianchi}, S. and 
                   {Camps}, P. and {Juvela}, M. and {Kuiper}, R.},
  title         = {{Composite biasing in Monte Carlo radiative transfer}},
  journal       = {\aap},
  archiveprefix = {arXiv},
  eprint        = {1603.07945},
  primaryclass  = {astro-ph.IM},
  year          = 2016,
  month         = may,
  volume        = 590,
  eid           = {A55},
  pages         = {A55},
  doi           = {10.1051/0004-6361/201528063},
  adsurl        = {http://adsabs.harvard.edu/abs/2016A%26A...590A..55B}
}

@article{2017A&C....20...16V,
  author        = {{Verstocken}, S. and {Van De Putte}, D. and {Camps}, P. and 
                   {Baes}, M.},
  title         = {{SKIRT: Hybrid parallelization of radiative transfer simulations}},
  journal       = {Astronomy and Computing},
  archiveprefix = {arXiv},
  eprint        = {1705.04702},
  primaryclass  = {astro-ph.IM},
  year          = 2017,
  month         = jul,
  volume        = 20,
  pages         = {16-33},
  doi           = {10.1016/j.ascom.2017.05.003},
  adsurl        = {http://adsabs.harvard.edu/abs/2017A%26C....20...16V}
}

@article{2013A&A...554A..10S,
  author        = {{Saftly}, W. and {Camps}, P. and {Baes}, M. and {Gordon}, K.~D. and 
                   {Vandewoude}, S. and {Rahimi}, A. and {Stalevski}, M.},
  title         = {{Using hierarchical octrees in Monte Carlo radiative transfer simulations}},
  journal       = {\aap},
  archiveprefix = {arXiv},
  eprint        = {1304.2896},
  primaryclass  = {astro-ph.IM},
  year          = 2013,
  month         = jun,
  volume        = 554,
  eid           = {A10},
  pages         = {A10},
  doi           = {10.1051/0004-6361/201220854},
  adsurl        = {http://adsabs.harvard.edu/abs/2013A%26A...554A..10S}
}

@article{2014A&A...561A..77S,
  author        = {{Saftly}, W. and {Baes}, M. and {Camps}, P.},
  title         = {{Hierarchical octree and k-d tree grids for 3D radiative transfer simulations}},
  journal       = {\aap},
  archiveprefix = {arXiv},
  eprint        = {1311.0705},
  primaryclass  = {astro-ph.IM},
  year          = 2014,
  month         = jan,
  volume        = 561,
  eid           = {A77},
  pages         = {A77},
  doi           = {10.1051/0004-6361/201322593},
  adsurl        = {http://adsabs.harvard.edu/abs/2014A%26A...561A..77S}
}

@article{2011ApJS..196...22B,
  author        = {{Baes}, M. and {Verstappen}, J. and {De Looze}, I. and {Fritz}, J. and 
                   {Saftly}, W. and {Vidal P{\'e}rez}, E. and {Stalevski}, M. and 
                   {Valcke}, S.},
  title         = {{Efficient Three-dimensional NLTE Dust Radiative Transfer with SKIRT}},
  journal       = {\apjs},
  archiveprefix = {arXiv},
  eprint        = {1108.5056},
  year          = 2011,
  month         = oct,
  volume        = 196,
  eid           = {22},
  pages         = {22},
  doi           = {10.1088/0067-0049/196/2/22},
  adsurl        = {http://adsabs.harvard.edu/abs/2011ApJS..196...22B}
}

@article{2015A&C.....9...20C,
  author        = {{Camps}, P. and {Baes}, M.},
  title         = {{SKIRT: An advanced dust radiative transfer code with a user-friendly architecture}},
  journal       = {Astronomy and Computing},
  archiveprefix = {arXiv},
  eprint        = {1410.1629},
  primaryclass  = {astro-ph.IM},
  year          = 2015,
  month         = mar,
  volume        = 9,
  pages         = {20-33},
  doi           = {10.1016/j.ascom.2014.10.004},
  adsurl        = {http://adsabs.harvard.edu/abs/2015A%26C.....9...20C}
}

@article{2013A&A...560A..35C,
  author        = {{Camps}, P. and {Baes}, M. and {Saftly}, W.},
  title         = {{Using 3D Voronoi grids in radiative transfer simulations}},
  journal       = {\aap},
  year          = 2013,
  month         = dec,
  volume        = {560},
  eid           = {A35},
  pages         = {A35},
  doi           = {10.1051/0004-6361/201322281},
  archiveprefix = {arXiv},
  eprint        = {1310.1854},
  primaryclass  = {astro-ph.IM},
  adsurl        = {https://ui.adsabs.harvard.edu/abs/2013A&A...560A..35C}
}

@article{2019MNRAS.489.4233M,
  author        = {{Marinacci}, Federico and {Sales}, Laura V. and {Vogelsberger}, Mark and {Torrey}, Paul and {Springel}, Volker},
  title         = {{Simulating the interstellar medium and stellar feedback on a moving mesh: implementation and isolated galaxies}},
  journal       = {\mnras},
  year          = 2019,
  month         = nov,
  volume        = {489},
  number        = {3},
  pages         = {4233-4260},
  doi           = {10.1093/mnras/stz2391},
  archiveprefix = {arXiv},
  eprint        = {1905.08806},
  primaryclass  = {astro-ph.GA},
  adsurl        = {https://ui.adsabs.harvard.edu/abs/2019MNRAS.489.4233M}
}

@article{2020A&C....3100381C,
  author        = {{Camps}, P. and {Baes}, M.},
  title         = {{SKIRT 9: Redesigning an advanced dust radiative transfer code to allow kinematics, line transfer and polarization by aligned dust grains}},
  journal       = {Astronomy and Computing},
  year          = 2020,
  month         = apr,
  volume        = {31},
  eid           = {100381},
  pages         = {100381},
  doi           = {10.1016/j.ascom.2020.100381},
  archiveprefix = {arXiv},
  eprint        = {2003.00721},
  primaryclass  = {astro-ph.GA},
  adsurl        = {https://ui.adsabs.harvard.edu/abs/2020A&C....3100381C}
}

@article{2010MNRAS.401..791S,
  author        = {{Springel}, Volker},
  title         = {{E pur si muove: Galilean-invariant cosmological hydrodynamical simulations on a moving mesh}},
  journal       = {\mnras},
  year          = 2010,
  month         = jan,
  volume        = {401},
  number        = {2},
  pages         = {791-851},
  doi           = {10.1111/j.1365-2966.2009.15715.x},
  archiveprefix = {arXiv},
  eprint        = {0901.4107},
  primaryclass  = {astro-ph.CO},
  adsurl        = {https://ui.adsabs.harvard.edu/abs/2010MNRAS.401..791S}
}

@article{2001ApJ...556..121K,
  author        = {{Kewley}, L.~J. and {Dopita}, M.~A. and {Sutherland}, R.~S. and {Heisler}, C.~A. and {Trevena}, J.},
  title         = {{Theoretical Modeling of Starburst Galaxies}},
  journal       = {\apj},
  year          = 2001,
  month         = jul,
  volume        = {556},
  number        = {1},
  pages         = {121-140},
  doi           = {10.1086/32154510.48550/arXiv.astro-ph/0106324},
  archiveprefix = {arXiv},
  eprint        = {astro-ph/0106324},
  primaryclass  = {astro-ph},
  adsurl        = {https://ui.adsabs.harvard.edu/abs/2001ApJ...556..121K}
}

@article{2019ARA&A..57..511K,
  author        = {{Kewley}, Lisa J. and {Nicholls}, David C. and {Sutherland}, Ralph S.},
  title         = {{Understanding Galaxy Evolution Through Emission Lines}},
  journal       = {\araa},
  year          = 2019,
  month         = aug,
  volume        = {57},
  pages         = {511-570},
  doi           = {10.1146/annurev-astro-081817-05183210.48550/arXiv.1910.09730},
  archiveprefix = {arXiv},
  eprint        = {1910.09730},
  primaryclass  = {astro-ph.GA},
  adsurl        = {https://ui.adsabs.harvard.edu/abs/2019ARA&A..57..511K}
}

@article{2003MNRAS.346.1055K,
  author        = {{Kauffmann}, Guinevere and {Heckman}, Timothy M. and {Tremonti}, Christy and {Brinchmann}, Jarle and {Charlot}, St{\'e}phane and {White}, Simon D.~M. and {Ridgway}, Susan E. and {Brinkmann}, Jon and {Fukugita}, Masataka and {Hall}, Patrick B. and {Ivezi{\'c}}, {\v{Z}}eljko and {Richards}, Gordon T. and {Schneider}, Donald P.},
  title         = {{The host galaxies of active galactic nuclei}},
  journal       = {\mnras},
  year          = 2003,
  month         = dec,
  volume        = {346},
  number        = {4},
  pages         = {1055-1077},
  doi           = {10.1111/j.1365-2966.2003.07154.x},
  archiveprefix = {arXiv},
  eprint        = {astro-ph/0304239},
  primaryclass  = {astro-ph},
  adsurl        = {https://ui.adsabs.harvard.edu/abs/2003MNRAS.346.1055K}
}

@article{2016MNRAS.461.3111B,
  author        = {{Belfiore}, Francesco and {Maiolino}, Roberto and {Maraston}, Claudia and {Emsellem}, Eric and {Bershady}, Matthew A. and {Masters}, Karen L. and {Yan}, Renbin and {Bizyaev}, Dmitry and {Boquien}, M{\'e}d{\'e}ric and {Brownstein}, Joel R. and {Bundy}, Kevin and {Drory}, Niv and {Heckman}, Timothy M. and {Law}, David R. and {Roman-Lopes}, Alexandre and {Pan}, Kaike and {Stanghellini}, Letizia and {Thomas}, Daniel and {Weijmans}, Anne-Marie and {Westfall}, Kyle B.},
  title         = {{SDSS IV MaNGA - spatially resolved diagnostic diagrams: a proof that many galaxies are LIERs}},
  journal       = {\mnras},
  year          = 2016,
  month         = sep,
  volume        = {461},
  number        = {3},
  pages         = {3111-3134},
  doi           = {10.1093/mnras/stw1234},
  archiveprefix = {arXiv},
  eprint        = {1605.07189},
  primaryclass  = {astro-ph.GA},
  adsurl        = {https://ui.adsabs.harvard.edu/abs/2016MNRAS.461.3111B}
}

@article{2019MNRAS.485..117K,
  author        = {{Kannan}, Rahul and {Vogelsberger}, Mark and {Marinacci}, Federico and {McKinnon}, Ryan and {Pakmor}, R{\"u}diger and {Springel}, Volker},
  title         = {{AREPO-RT: radiation hydrodynamics on a moving mesh}},
  journal       = {\mnras},
  year          = 2019,
  month         = may,
  volume        = {485},
  number        = {1},
  pages         = {117-149},
  doi           = {10.1093/mnras/stz287},
  archiveprefix = {arXiv},
  eprint        = {1804.01987},
  primaryclass  = {astro-ph.IM},
  adsurl        = {https://ui.adsabs.harvard.edu/abs/2019MNRAS.485..117K}
}

@article{2020ApJ...905...27S,
  author        = {{Smith}, Aaron and {Kannan}, Rahul and {Tsang}, Benny T. -H. and {Vogelsberger}, Mark and {Pakmor}, R{\"u}diger},
  title         = {{AREPO-MCRT: Monte Carlo Radiation Hydrodynamics on a Moving Mesh}},
  journal       = {\apj},
  year          = 2020,
  month         = dec,
  volume        = {905},
  number        = {1},
  eid           = {27},
  pages         = {27},
  doi           = {10.3847/1538-4357/abc47e},
  archiveprefix = {arXiv},
  eprint        = {2008.01750},
  primaryclass  = {astro-ph.GA},
  adsurl        = {https://ui.adsabs.harvard.edu/abs/2020ApJ...905...27S}
}

@article{2021ApJ...916...39C,
  author        = {{Camps}, Peter and {Behrens}, Christoph and {Baes}, Maarten and {Kapoor}, Anand Utsav and {Grand}, Robert},
  title         = {{Effects of Spatial Discretization in Ly{\ensuremath{\alpha}} Line Radiation Transfer Simulations}},
  journal       = {\apj},
  year          = 2021,
  month         = jul,
  volume        = {916},
  number        = {1},
  eid           = {39},
  pages         = {39},
  doi           = {10.3847/1538-4357/ac06cb},
  archiveprefix = {arXiv},
  eprint        = {2106.00281},
  primaryclass  = {astro-ph.IM},
  adsurl        = {https://ui.adsabs.harvard.edu/abs/2021ApJ...916...39C}
}

@article{2022MNRAS.517....1S,
  author        = {{Smith}, Aaron and {Kannan}, Rahul and {Tacchella}, Sandro and {Vogelsberger}, Mark and {Hernquist}, Lars and {Marinacci}, Federico and {Sales}, Laura V. and {Torrey}, Paul and {Li}, Hui and {Yeh}, Jessica Y. -C. and {Qi}, Jia},
  title         = {{The physics of Lyman-{\ensuremath{\alpha}} escape from disc-like galaxies}},
  journal       = {\mnras},
  year          = 2022,
  month         = nov,
  volume        = {517},
  number        = {1},
  pages         = {1-27},
  doi           = {10.1093/mnras/stac2641},
  archiveprefix = {arXiv},
  eprint        = {2111.13721},
  primaryclass  = {astro-ph.GA},
  adsurl        = {https://ui.adsabs.harvard.edu/abs/2022MNRAS.517....1S}
}

@article{2022MNRAS.513.2904T,
  author        = {{Tacchella}, Sandro and {Smith}, Aaron and {Kannan}, Rahul and {Marinacci}, Federico and {Hernquist}, Lars and {Vogelsberger}, Mark and {Torrey}, Paul and {Sales}, Laura and {Li}, Hui},
  title         = {{H {\ensuremath{\alpha}} emission in local galaxies: star formation, time variability, and the diffuse ionized gas}},
  journal       = {\mnras},
  year          = 2022,
  month         = jun,
  volume        = {513},
  number        = {2},
  pages         = {2904-2929},
  doi           = {10.1093/mnras/stac818},
  archiveprefix = {arXiv},
  eprint        = {2112.00027},
  primaryclass  = {astro-ph.GA},
  adsurl        = {https://ui.adsabs.harvard.edu/abs/2022MNRAS.513.2904T}
}

@article{2021A&A...653A..34V,
  author        = {{Vandenbroucke}, B. and {Baes}, M. and {Camps}, P. and {Kapoor}, A.~U. and {Barrientos}, D. and {Bernard}, J. -P.},
  title         = {{Polarised emission from aligned dust grains in nearby galaxies: Predictions from the Auriga simulations}},
  journal       = {\aap},
  year          = 2021,
  month         = sep,
  volume        = {653},
  eid           = {A34},
  pages         = {A34},
  doi           = {10.1051/0004-6361/202141333},
  archiveprefix = {arXiv},
  eprint        = {2107.02180},
  primaryclass  = {astro-ph.GA},
  adsurl        = {https://ui.adsabs.harvard.edu/abs/2021A&A...653A..34V}
}

@article{2015A&A...580A..87C,
  author        = {{Camps}, Peter and {Misselt}, Karl and {Bianchi}, Simone and {Lunttila}, Tuomas and {Pinte}, Christophe and {Natale}, Giovanni and {Juvela}, Mika and {Fischera}, Joerg and {Fitzgerald}, Michael P. and {Gordon}, Karl and {Baes}, Maarten and {Steinacker}, J{\"u}rgen},
  title         = {{Benchmarking the calculation of stochastic heating and emissivity of dust grains in the context of radiative transfer simulations}},
  journal       = {\aap},
  year          = 2015,
  month         = aug,
  volume        = {580},
  eid           = {A87},
  pages         = {A87},
  doi           = {10.1051/0004-6361/201525998},
  archiveprefix = {arXiv},
  eprint        = {1506.05304},
  primaryclass  = {astro-ph.IM},
  adsurl        = {https://ui.adsabs.harvard.edu/abs/2015A&A...580A..87C}
}

@article{2023MNRAS.526.3871K,
  author        = {{Kapoor}, Anand Utsav and {Baes}, Maarten and {van der Wel}, Arjen and {Gebek}, Andrea and {Camps}, Peter and {Nersesian}, Angelos and {Meidt}, Sharon E. and {Smith}, Aaron and {Vicens}, Sebastien and {D'Eugenio}, Francesco and {Martorano}, Marco and {Barrientos}, Daniela and {Sartorio}, Nina Sanches},
  title         = {{TODDLERS: a new UV-mm emission library for star-forming regions - I. Integration with SKIRT and public release}},
  journal       = {\mnras},
  year          = 2023,
  month         = dec,
  volume        = {526},
  number        = {3},
  pages         = {3871-3901},
  doi           = {10.1093/mnras/stad2977},
  archiveprefix = {arXiv},
  eprint        = {2310.00388},
  primaryclass  = {astro-ph.GA},
  adsurl        = {https://ui.adsabs.harvard.edu/abs/2023MNRAS.526.3871K}
}

@article{2017PASA...34...58E,
  author        = {{Eldridge}, J.~J. and {Stanway}, E.~R. and {Xiao}, L. and {McClelland}, L.~A.~S. and {Taylor}, G. and {Ng}, M. and {Greis}, S.~M.~L. and {Bray}, J.~C.},
  title         = {{Binary Population and Spectral Synthesis Version 2.1: Construction, Observational Verification, and New Results}},
  journal       = {\pasa},
  year          = 2017,
  month         = nov,
  volume        = {34},
  eid           = {e058},
  pages         = {e058},
  doi           = {10.1017/pasa.2017.51},
  archiveprefix = {arXiv},
  eprint        = {1710.02154},
  primaryclass  = {astro-ph.SR},
  adsurl        = {https://ui.adsabs.harvard.edu/abs/2017PASA...34...58E}
}

@ARTICLE{2022A&A...659A..26B,
       author = {{Belfiore}, F. and {Santoro}, F. and {Groves}, B. and {Schinnerer}, E. and {Kreckel}, K. and {Glover}, S.~C.~O. and {Klessen}, R.~S. and {Emsellem}, E. and {Blanc}, G.~A. and {Congiu}, E. and {Barnes}, A.~T. and {Boquien}, M. and {Chevance}, M. and {Dale}, D.~A. and {Kruijssen}, J.~M. Diederik and {Leroy}, A.~K. and {Pan}, H.-A. and {Pessa}, I. and {Schruba}, A. and {Williams}, T.~G.},
        title = "{A tale of two DIGs: The relative role of H II regions and low-mass hot evolved stars in powering the diffuse ionised gas (DIG) in PHANGS-MUSE galaxies}",
      journal = {\aap},
         year = 2022,
        month = mar,
       volume = {659},
          eid = {A26},
        pages = {A26},
          doi = {10.1051/0004-6361/202141859},
archivePrefix = {arXiv},
       eprint = {2111.14876},
 primaryClass = {astro-ph.GA},
       adsurl = {https://ui.adsabs.harvard.edu/abs/2022A&A...659A..26B}
}

@ARTICLE{2025arXiv251013952M,
       author = {{McClymont}, William and {Smith}, Aaron and {Tacchella}, Sandro},
        title = "{Modelling the nebular emission of galaxies across cosmic time with COLT}",
      journal = {arXiv e-prints},
         year = 2025,
        month = oct,
          eid = {arXiv:2510.13952},
        pages = {arXiv:2510.13952},
          doi = {10.48550/arXiv.2510.13952},
archivePrefix = {arXiv},
       eprint = {2510.13952},
 primaryClass = {astro-ph.GA},
       adsurl = {https://ui.adsabs.harvard.edu/abs/2025arXiv251013952M}
}

@ARTICLE{2024A&A...692A..79K,
       author = {{Kapoor}, Anand Utsav and {Baes}, Maarten and {van der Wel}, Arjen and {Gebek}, Andrea and {Camps}, Peter and {Smith}, Aaron and {Boquien}, M{\'e}d{\'e}ric and {Andreadis}, Nick and {Vicens}, Sebastien},
        title = "{TODDLERS: A new UV-millimeter emission library for star-forming regions: II. Star-formation rate indicators using Auriga zoom simulations}",
      journal = {\aap},
         year = 2024,
        month = dec,
       volume = {692},
          eid = {A79},
        pages = {A79},
          doi = {10.1051/0004-6361/202451207},
archivePrefix = {arXiv},
       eprint = {2410.01067},
 primaryClass = {astro-ph.GA},
       adsurl = {https://ui.adsabs.harvard.edu/abs/2024A&A...692A..79K}
}

@ARTICLE{2024MNRAS.532.2016M,
       author = {{McClymont}, William and {Tacchella}, Sandro and {Smith}, Aaron and {Kannan}, Rahul and {Maiolino}, Roberto and {Belfiore}, Francesco and {Hernquist}, Lars and {Li}, Hui and {Vogelsberger}, Mark},
        title = "{The nature of diffuse ionized gas in star-forming galaxies}",
      journal = {\mnras},
         year = 2024,
        month = aug,
       volume = {532},
       number = {2},
        pages = {2016-2031},
          doi = {10.1093/mnras/stae1587},
archivePrefix = {arXiv},
       eprint = {2403.03243},
 primaryClass = {astro-ph.GA},
       adsurl = {https://ui.adsabs.harvard.edu/abs/2024MNRAS.532.2016M}
}

@ARTICLE{2020MNRAS.499.5732K,
       author = {{Kannan}, Rahul and {Marinacci}, Federico and {Vogelsberger}, Mark and {Sales}, Laura V. and {Torrey}, Paul and {Springel}, Volker and {Hernquist}, Lars},
        title = "{Simulating the interstellar medium of galaxies with radiative transfer, non-equilibrium thermochemistry, and dust}",
      journal = {\mnras},
         year = 2020,
        month = dec,
       volume = {499},
       number = {4},
        pages = {5732-5748},
          doi = {10.1093/mnras/staa3249},
archivePrefix = {arXiv},
       eprint = {1910.14041},
 primaryClass = {astro-ph.GA},
       adsurl = {https://ui.adsabs.harvard.edu/abs/2020MNRAS.499.5732K}
}

@ARTICLE{2020ARA&A..58..661F,
       author = {{F{\"o}rster Schreiber}, Natascha M. and {Wuyts}, Stijn},
        title = "{Star-Forming Galaxies at Cosmic Noon}",
      journal = {\araa},
         year = 2020,
        month = aug,
       volume = {58},
        pages = {661-725},
          doi = {10.1146/annurev-astro-032620-021910},
archivePrefix = {arXiv},
       eprint = {2010.10171},
 primaryClass = {astro-ph.GA},
       adsurl = {https://ui.adsabs.harvard.edu/abs/2020ARA&A..58..661F}
}

@ARTICLE{2018A&C....23...40V,
       author = {{Vandenbroucke}, B. and {Wood}, K.},
        title = "{The Monte Carlo photoionization and moving-mesh radiation hydrodynamics code CMACIONIZE}",
      journal = {Astronomy and Computing},
         year = 2018,
        month = apr,
       volume = {23},
          eid = {40},
        pages = {40},
          doi = {10.1016/j.ascom.2018.02.005},
       adsurl = {https://ui.adsabs.harvard.edu/abs/2018A&C....23...40V}
}

@ARTICLE{2004MNRAS.348.1337W,
       author = {{Wood}, Kenneth and {Mathis}, John S. and {Ercolano}, Barbara},
        title = "{A three-dimensional Monte Carlo photoionization code for modelling diffuse ionized gas}",
      journal = {\mnras},
         year = 2004,
        month = mar,
       volume = {348},
       number = {4},
        pages = {1337-1347},
          doi = {10.1111/j.1365-2966.2004.07458.x},
archivePrefix = {arXiv},
       eprint = {astro-ph/0311584},
 primaryClass = {astro-ph},
       adsurl = {https://ui.adsabs.harvard.edu/abs/2004MNRAS.348.1337W}
}

@ARTICLE{1996ApJ...465..487V,
       author = {{Verner}, D.~A. and {Ferland}, G.~J. and {Korista}, K.~T. and {Yakovlev}, D.~G.},
        title = "{Atomic Data for Astrophysics. II. New Analytic Fits for Photoionization Cross Sections of Atoms and Ions}",
      journal = {\apj},
         year = 1996,
        month = jul,
       volume = {465},
        pages = {487},
          doi = {10.1086/177435},
archivePrefix = {arXiv},
       eprint = {astro-ph/9601009},
 primaryClass = {astro-ph},
       adsurl = {https://ui.adsabs.harvard.edu/abs/1996ApJ...465..487V}
}

@ARTICLE{1996ApJS..103..467V,
       author = {{Verner}, D.~A. and {Ferland}, G.~J.},
        title = "{Atomic Data for Astrophysics. I. Radiative Recombination Rates for H-like, He-like, Li-like, and Na-like Ions over a Broad Range of Temperature}",
      journal = {\apjs},
         year = 1996,
        month = apr,
       volume = {103},
        pages = {467},
          doi = {10.1086/192284},
archivePrefix = {arXiv},
       eprint = {astro-ph/9509083},
 primaryClass = {astro-ph},
       adsurl = {https://ui.adsabs.harvard.edu/abs/1996ApJS..103..467V}
}

@ARTICLE{1969PhRv..180...25D,
       author = {{Drake}, G.~W. and {Victor}, G.~A. and {Dalgarno}, A.},
        title = "{Two-Photon Decay of the Singlet and Triplet Metastable States of Helium-like Ions}",
      journal = {Physical Review},
         year = 1969,
        month = apr,
       volume = {180},
       number = {1},
        pages = {25-32},
          doi = {10.1103/PhysRev.180.25},
       adsurl = {https://ui.adsabs.harvard.edu/abs/1969PhRv..180...25D}
}

@misc{2018ascl.soft07005S,
       author = {{Sutherland}, Ralph and {Dopita}, Mike and {Binette}, Luc and {Groves}, Brent},
        title = "{MAPPINGS V: Astrophysical plasma modeling code}",
 howpublished = {Astrophysics Source Code Library, record ascl:1807.005},
         year = 2018,
        month = jul,
          eid = {ascl:1807.005},
archivePrefix = {ascl},
       eprint = {1807.005},
       adsurl = {https://ui.adsabs.harvard.edu/abs/2018ascl.soft07005S}
}

@ARTICLE{1997MNRAS.292...27H,
       author = {{Hui}, Lam and {Gnedin}, Nickolay Y.},
        title = "{Equation of state of the photoionized intergalactic medium}",
      journal = {\mnras},
         year = 1997,
        month = nov,
       volume = {292},
       number = {1},
        pages = {27-42},
          doi = {10.1093/mnras/292.1.27},
archivePrefix = {arXiv},
       eprint = {astro-ph/9612232},
 primaryClass = {astro-ph},
       adsurl = {https://ui.adsabs.harvard.edu/abs/1997MNRAS.292...27H}
}

@ARTICLE{2003MNRAS.340.1136E,
       author = {{Ercolano}, B. and {Barlow}, M.~J. and {Storey}, P.~J. and {Liu}, X.-W.},
        title = "{MOCASSIN: a fully three-dimensional Monte Carlo photoionization code}",
      journal = {\mnras},
         year = 2003,
        month = apr,
       volume = {340},
       number = {4},
        pages = {1136-1152},
          doi = {10.1046/j.1365-8711.2003.06371.x},
archivePrefix = {arXiv},
       eprint = {astro-ph/0209378},
 primaryClass = {astro-ph},
       adsurl = {https://ui.adsabs.harvard.edu/abs/2003MNRAS.340.1136E}
}

@ARTICLE{2005MNRAS.362.1038E,
       author = {{Ercolano}, B. and {Barlow}, M.~J. and {Storey}, P.~J.},
        title = "{The dusty MOCASSIN: fully self-consistent 3D photoionization and dust radiative transfer models}",
      journal = {\mnras},
         year = 2005,
        month = sep,
       volume = {362},
       number = {3},
        pages = {1038-1046},
          doi = {10.1111/j.1365-2966.2005.09381.x},
archivePrefix = {arXiv},
       eprint = {astro-ph/0507050},
 primaryClass = {astro-ph},
       adsurl = {https://ui.adsabs.harvard.edu/abs/2005MNRAS.362.1038E}
}

@ARTICLE{2018MNRAS.479..994R,
       author = {{Rosdahl}, Joakim and {Katz}, Harley and {Blaizot}, J{\'e}r{\'e}mie and {Kimm}, Taysun and {Michel-Dansac}, L{\'e}o and {Garel}, Thibault and {Haehnelt}, Martin and {Ocvirk}, Pierre and {Teyssier}, Romain},
        title = "{The SPHINX cosmological simulations of the first billion years: the impact of binary stars on reionization}",
      journal = {\mnras},
         year = 2018,
        month = sep,
       volume = {479},
       number = {1},
        pages = {994--1016},
          doi = {10.1093/mnras/sty1655},
archivePrefix = {arXiv},
       eprint = {1801.07259},
 primaryClass = {astro-ph.GA},
       adsurl = {https://ui.adsabs.harvard.edu/abs/2018MNRAS.479..994R}
}

@ARTICLE{2023OJAp....6E..44K,
       author = {{Katz}, Harley and {Rosdahl}, Joki and {Kimm}, Taysun and {Blaizot}, Jeremy and {Choustikov}, Nicholas and {Farcy}, Marion and {Garel}, Thibault and {Haehnelt}, Martin G. and {Michel-Dansac}, Leo and {Ocvirk}, Pierre},
        title = "{The SPHINX Public Data Release: Forward Modelling High-Redshift JWST Observations with Cosmological Radiation Hydrodynamics Simulations}",
      journal = {The Open Journal of Astrophysics},
         year = 2023,
        month = dec,
       volume = {6},
          eid = {44},
        pages = {44},
          doi = {10.21105/astro.2309.03269},
archivePrefix = {arXiv},
       eprint = {2309.03269},
 primaryClass = {astro-ph.GA},
       adsurl = {https://ui.adsabs.harvard.edu/abs/2023OJAp....6E..44K}
}

@ARTICLE{2022MNRAS.511.4005K,
       author = {{Kannan}, R. and {Garaldi}, E. and {Smith}, A. and {Pakmor}, R. and {Springel}, V. and {Vogelsberger}, M. and {Hernquist}, L.},
        title = "{Introducing the THESAN project: radiation-magnetohydrodynamic simulations of the epoch of reionization}",
      journal = {\mnras},
         year = 2022,
       volume = {511},
       number = {3},
        pages = {4005-4030},
          doi = {10.1093/mnras/stab3710},
       eprint = {2110.00584},
}

@ARTICLE{2025OJAp....8E.153K,
       author = {{Kannan}, Rahul and {Puchwein}, Ewald and {Smith}, Aaron and {Borrow}, Josh and {Garaldi}, Enrico and {Keating}, Laura and {Vogelsberger}, Mark and {Zier}, Oliver and {McClymont}, William and {Shen}, Xuejian and {Popovic}, Filip and {Tacchella}, Sandro and {Hernquist}, Lars and {Springel}, Volker},
        title = "{Introducing the THESAN-ZOOM project: radiation-hydrodynamic simulations of high-redshift galaxies with a multi-phase interstellar medium}",
      journal = {The Open Journal of Astrophysics},
         year = 2025,
       volume = {8},
          eid = {153},
        pages = {153},
          doi = {10.33232/001c.145804},
       eprint = {2502.20437},
}

@ARTICLE{2025arXiv251005201K,
       author = {{Katz}, Harley and {Rey}, Martin P. and {Cadiou}, Corentin and {Agertz}, Oscar and {Blaizot}, Jeremy and {Cameron}, Alex J. and {Choustikov}, Nicholas and {Devriendt}, Julien and {Hauk}, Uliana and {Jones}, Gareth C. and {Kimm}, Taysun and {Laseter}, Isaac and {Martin-Alvarez}, Sergio and {Matsumoto}, Kosei and {Pearce}, Autumn and {Rodr{\'{\i}}guez Montero}, Francisco and {Rosdahl}, Joki and {Sanati}, Mahsa and {Saxena}, Aayush and {Slyz}, Adrianne and {Stiskalek}, Richard and {Storck}, Anatole and {Veenema}, Oscar and {Yee}, Wonjae},
        title = "{MEGATRON: Reproducing the Diversity of High-Redshift Galaxy Spectra with Cosmological Radiation Hydrodynamics Simulations}",
      journal = {arXiv e-prints},
         year = 2025,
          eid = {arXiv:2510.05201},
        pages = {arXiv:2510.05201},
          doi = {10.48550/arXiv.2510.05201},
       eprint = {2510.05201},
}

@ARTICLE{2012MNRAS.424..884Y,
       author = {{Yajima}, Hidenobu and {Li}, Yuexing and {Zhu}, Qirong and {Abel}, Tom},
        title = "{ART$^{2}$: coupling Ly{\ensuremath{\alpha}} line and multi-wavelength continuum radiative transfer}",
      journal = {\mnras},
         year = 2012,
       volume = {424},
       number = {2},
        pages = {884-901},
          doi = {10.1111/j.1365-2966.2012.21228.x},
       eprint = {1109.4891},
}

@ARTICLE{2020MNRAS.494.1919L,
       author = {{Li}, Yuexing and {Gu}, Ming F. and {Yajima}, Hidenobu and {Zhu}, Qirong and {Maji}, Moupiya},
        title = "{ART$^{2}$: a 3D parallel multiwavelength radiative transfer code for continuum and atomic and molecular lines}",
      journal = {\mnras},
         year = 2020,
       volume = {494},
       number = {2},
        pages = {1919-1935},
          doi = {10.1093/mnras/staa733},
       eprint = {2001.11146},
}

@ARTICLE{1997A&AS..125..149D,
       author = {{Dere}, K.~P. and {Landi}, E. and {Mason}, H.~E. and {Monsignori Fossi}, B.~C. and {Young}, P.~R.},
        title = "{CHIANTI - an atomic database for emission lines}",
      journal = {\aaps},
         year = 1997,
       volume = {125},
        pages = {149-173},
          doi = {10.1051/aas:1997368},
}

@ARTICLE{2021ApJ...909...38D,
       author = {{Del Zanna}, G. and {Dere}, K.~P. and {Young}, P.~R. and {Landi}, E.},
        title = "{CHIANTI{\textemdash}An Atomic Database for Emission Lines. XVI. Version 10, Further Extensions}",
      journal = {\apj},
         year = 2021,
       volume = {909},
       number = {1},
          eid = {38},
        pages = {38},
          doi = {10.3847/1538-4357/abd8ce},
       eprint = {2011.05211},
}

@ARTICLE{1995MNRAS.272...41S,
       author = {{Storey}, P.~J. and {Hummer}, D.~G.},
        title = "{Recombination line intensities for hydrogenic ions-IV. Total recombination coefficients and machine-readable tables for Z=1 to 8}",
      journal = {\mnras},
         year = 1995,
       volume = {272},
       number = {1},
        pages = {41-48},
          doi = {10.1093/mnras/272.1.41},
}

@ARTICLE{2006MNRAS.372.1875E,
       author = {{Ercolano}, B. and {Storey}, P.~J.},
        title = "{Theoretical calculations of the H I, He I and He II free-bound continuous emission spectra}",
      journal = {\mnras},
         year = 2006,
       volume = {372},
       number = {4},
        pages = {1875-1878},
          doi = {10.1111/j.1365-2966.2006.10988.x},
       eprint = {astro-ph/0609174},
}

@ARTICLE{2025arXiv250821126S,
       author = {{Schaye}, Joop and {Chaikin}, Evgenii and {Schaller}, Matthieu and {Ploeckinger}, Sylvia and {Hu{\v{s}}ko}, Filip and {McGibbon}, Rob and {Trayford}, James W. and {Ben{\'{\i}}tez-Llambay}, Alejandro and {Correa}, Camila and {Frenk}, Carlos S. and {Richings}, Alexander J. and {Forouhar Moreno}, Victor J. and {Bah{\'e}}, Yannick M. and {Borrow}, Josh and {Durrant}, Anna and {Gebek}, Andrea and {Helly}, John C. and {Jenkins}, Adrian and {Lacey}, Cedric G. and {Ludlow}, Aaron and {Nobels}, Folkert S.~J.},
        title = "{The COLIBRE project: cosmological hydrodynamical simulations of galaxy formation and evolution}",
      journal = {arXiv e-prints},
         year = 2025,
          eid = {arXiv:2508.21126},
        pages = {arXiv:2508.21126},
          doi = {10.48550/arXiv.2508.21126},
       eprint = {2508.21126},
}

@ARTICLE{2025ApJ...980..242S,
       author = {{Shapley}, Alice E. and {Sanders}, Ryan L. and {Topping}, Michael W. and {Reddy}, Naveen A. and {Berg}, Danielle A. and {Bouwens}, Rychard J. and {Brammer}, Gabriel and {Carnall}, Adam C. and {Cullen}, Fergus and {Dav{\'e}}, Romeel and {Dunlop}, James S. and {Ellis}, Richard S. and {F{\"o}rster Schreiber}, N.~M. and {Furlanetto}, Steven R. and {Glazebrook}, Karl and {Illingworth}, Garth D. and {Jones}, Tucker and {Kriek}, Mariska and {McLeod}, Derek J. and {McLure}, Ross J. and {Narayanan}, Desika and {Oesch}, Pascal and {Pahl}, Anthony J. and {Pettini}, Max and {Schaerer}, Daniel and {Stark}, Daniel P. and {Steidel}, Charles C. and {Tang}, Mengtao and {Clarke}, Leonardo and {Donnan}, Callum T. and {Kehoe}, Emily},
        title = "{The AURORA Survey: A New Era of Emission-line Diagrams with JWST/NIRSpec}",
      journal = {\apj},
         year = 2025,
       volume = {980},
       number = {2},
          eid = {242},
        pages = {242},
          doi = {10.3847/1538-4357/adad68},
       eprint = {2407.00157},
}

@ARTICLE{2009RvMP...81..969H,
       author = {{Haffner}, L.~M. and {Dettmar}, R.-J. and {Beckman}, J.~E. and {Wood}, K. and {Slavin}, J.~D. and {Giammanco}, C. and {Madsen}, G.~J. and {Zurita}, A. and {Reynolds}, R.~J.},
        title = "{The warm ionized medium in spiral galaxies}",
      journal = {Reviews of Modern Physics},
         year = 2009,
       volume = {81},
       number = {3},
        pages = {969-997},
          doi = {10.1103/RevModPhys.81.969},
       eprint = {0901.0941},
}

@ARTICLE{2024A&A...689A.297V,
       author = {{Vander Meulen}, Bert and {Camps}, Peter and {Savi{\'c}}, {\DH}or{\dj}e and {Baes}, Maarten and {Matt}, Giorgio and {Stalevski}, Marko},
        title = "{X-ray polarisation in AGN circumnuclear media: Polarisation framework and 2D torus models}",
      journal = {\aap},
         year = 2024,
        month = sep,
       volume = {689},
          eid = {A297},
        pages = {A297},
          doi = {10.1051/0004-6361/202450773},
       adsurl = {https://ui.adsabs.harvard.edu/abs/2024A&A...689A.297V}
}

@ARTICLE{2022A&A...666A.101B,
       author = {{Baes}, Maarten and {Camps}, Peter and {Matsumoto}, Kosei},
        title = "{Monte Carlo radiative transfer with explicit absorption to simulate absorption, scattering, and stimulated emission}",
      journal = {\aap},
         year = 2022,
        month = oct,
       volume = {666},
          eid = {A101},
        pages = {A101},
          doi = {10.1051/0004-6361/202244521},
       adsurl = {https://ui.adsabs.harvard.edu/abs/2022A&A...666A.101B}
}

@ARTICLE{2025MNRAS.536..879M,
       author = {{Matthews}, James H. and {Long}, Knox S. and {Knigge}, Christian and {Sim}, Stuart A. and {Parkinson}, Edward J. and {Higginbottom}, Nick and {Mangham}, Samuel W. and {Scepi}, Nicolas and {Wallis}, Austen and {Hewitt}, Henrietta A. and {Mosallanezhad}, Amin},
        title = "{SIROCCO: a publicly available Monte Carlo ionization and radiative transfer code for astrophysical outflows}",
      journal = {\mnras},
         year = 2025,
        month = jan,
       volume = {536},
       number = {1},
        pages = {879-904},
          doi = {10.1093/mnras/stae2677},
       adsurl = {https://ui.adsabs.harvard.edu/abs/2025MNRAS.536..879M}
}

@ARTICLE{Smith2015,
       author = {{Smith}, Aaron and {Safranek-Shrader}, Chalence and {Bromm}, Volker and {Milosavljevi{\'c}}, Milo{\v{s}}},
        title = "{The Lyman {\ensuremath{\alpha}} signature of the first galaxies}",
      journal = {\mnras},
         year = 2015,
        month = jun,
       volume = {449},
       number = {4},
        pages = {4336-4362},
          doi = {10.1093/mnras/stv565},
archivePrefix = {arXiv},
       eprint = {1409.4480},
 primaryClass = {astro-ph.CO},
       adsurl = {https://ui.adsabs.harvard.edu/abs/2015MNRAS.449.4336S}
}

@ARTICLE{Smith2019,
       author = {{Smith}, Aaron and {Ma}, Xiangcheng and {Bromm}, Volker and {Finkelstein}, Steven L. and {Hopkins}, Philip F. and {Faucher-Gigu{\`e}re}, Claude-Andr{\'e} and {Kere{\v{s}}}, Du{\v{s}}an},
        title = "{The physics of Lyman {\ensuremath{\alpha}} escape from high-redshift galaxies}",
      journal = {\mnras},
         year = 2019,
        month = mar,
       volume = {484},
       number = {1},
        pages = {39-59},
          doi = {10.1093/mnras/sty3483},
archivePrefix = {arXiv},
       eprint = {1810.08185},
 primaryClass = {astro-ph.GA},
       adsurl = {https://ui.adsabs.harvard.edu/abs/2019MNRAS.484...39S}
}

@ARTICLE{Vandenbroucke2020,
       author = {{Vandenbroucke}, B. and {Camps}, P.},
        title = "{CMACIONIZE 2.0: a novel task-based approach to Monte Carlo radiation transfer}",
      journal = {\aap},
         year = 2020,
        month = sep,
       volume = {641},
          eid = {A66},
        pages = {A66},
          doi = {10.1051/0004-6361/202038364},
archivePrefix = {arXiv},
       eprint = {2006.15147},
 primaryClass = {astro-ph.IM},
       adsurl = {https://ui.adsabs.harvard.edu/abs/2020A&A...641A..66V}
}

@ARTICLE{McCallum2024,
       author = {{McCallum}, Lewis and {Wood}, Kenneth and {Benjamin}, Robert and {Krishnarao}, Dhanesh and {Vandenbroucke}, Bert},
        title = "{Time-dependent metal ionization and the persistence of collisionally excited emission lines in the diffuse ionized gas of star-forming galaxies}",
      journal = {\mnras},
         year = 2024,
        month = dec,
       volume = {535},
       number = {3},
        pages = {2889-2902},
          doi = {10.1093/mnras/stae2549},
archivePrefix = {arXiv},
       eprint = {2411.07108},
 primaryClass = {astro-ph.GA},
       adsurl = {https://ui.adsabs.harvard.edu/abs/2024MNRAS.535.2889M}
}

@ARTICLE{1994ApJ...427..822B,
       author = {{Bakes}, E.~L.~O. and {Tielens}, A.~G.~G.~M.},
        title = "{The Photoelectric Heating Mechanism for Very Small Graphitic Grains and Polycyclic Aromatic Hydrocarbons}",
      journal = {\apj},
         year = 1994,
        month = jun,
       volume = {427},
        pages = {822},
          doi = {10.1086/174188},
       adsurl = {https://ui.adsabs.harvard.edu/abs/1994ApJ...427..822B}
}

@ARTICLE{2001ApJS..134..263W,
       author = {{Weingartner}, Joseph C. and {Draine}, B.~T.},
        title = "{Photoelectric Emission from Interstellar Dust: Grain Charging and Gas Heating}",
      journal = {\apjs},
         year = 2001,
        month = jun,
       volume = {134},
       number = {2},
        pages = {263-281},
          doi = {10.1086/320852},
archivePrefix = {arXiv},
       eprint = {astro-ph/9907251},
 primaryClass = {astro-ph},
       adsurl = {https://ui.adsabs.harvard.edu/abs/2001ApJS..134..263W}
}

@ARTICLE{1983ApJ...265..223B,
       author = {{Burke}, J.~R. and {Hollenbach}, D.~J.},
        title = "{The gas-grain interaction in the interstellar medium - Thermal accommodation and trapping}",
      journal = {\apj},
         year = 1983,
        month = feb,
       volume = {265},
        pages = {223-234},
          doi = {10.1086/160667},
       adsurl = {https://ui.adsabs.harvard.edu/abs/1983ApJ...265..223B}
}
